\documentclass[useAMS,usenatbib]{mn2e} 
\usepackage{times,array,graphicx,amssymb,amsmath}

\usepackage{color}

\newcommand{\sfr}{\textrm{SFR}}
\newcommand{\pyr}{\textrm{yr}^{-1}}
\newcommand{\msun}{\rm M_\odot}
\newcommand{\vl}{{\mathbf{L}}}

\newcommand{\mha}{\textrm{H}\alpha }

\newcommand{\qho}{\textrm{Q(H}^0)}

\newcommand{\slug}{\texttt{SLUG}}
\newcommand{\slugt}{\texttt{SLUG}}
\newcommand{\sfrqho}{\sfr_{Q(H^0)}}
\newcommand{\sfrfuv}{\sfr_{\rm FUV}}
\newcommand{\sfrbol}{\sfr_{\rm BOL}}
\newcommand{\vizloc}{https://sites.google.com/site/runslug}

\input{journal.def}

\title[Stochastically Lighting Up Galaxies II]{SLUG -- Stochastically Lighting Up Galaxies. II: Quantifying the Effects of Stochasticity on Star Formation Rate Indicators}

\author[da Silva et al.]{Robert L. da Silva$^{1}$, Michele Fumagalli$^{2,3}$\thanks{E-mail: michele.fumagalli@durham.ac.uk}, and Mark R. Krumholz$^{1}$\thanks{mkrumhol@ucsc.edu}\\
$^{1}$Department of Astronomy and Astrophysics, University of California, 1156 High Street, Santa Cruz, CA 95064, USA \\
$^{2}$Institute for Computational Cosmology, Department of Physics, Durham University, South Road, Durham, DH1 3LE, UK \\
$^{3}$Carnegie Observatories, 813 Santa Barbara Street, Pasadena, CA 91101, USA \\}

\begin{document}

\date{Accepted xxxx. Received xxxx; in original form xxxx}

\pagerange{\pageref{firstpage}--\pageref{lastpage}} \pubyear{xxxx}

\maketitle

\label{firstpage}

\begin{abstract} 
The integrated light of a stellar population, measured through photometric filters that are sensitive to the presence of 
young stars, is often used to infer the star formation rate (SFR) for that population. However, these techniques rely on an 
assumption that star formation is a continuous process, whereas in reality stars form in discrete spatially- and 
temporally-correlated structures. This discreteness causes the light output to undergo significant time-dependent fluctuations, 
which, if not accounted for, introduce systematic errors in the inferred SFRs due to the intrinsic distribution of 
luminosities at any fix SFR. We use \slug\, a code that Stochastically 
Lights Up Galaxies, to simulate galaxies undergoing stochastic star formation. We then use these simulations to present a 
quantitative analysis of these effects and provide tools for calculating probability distribution functions of SFRs given 
a set of observations. We show that, depending on the SFR tracer used, stochastic fluctuations can produce non-trivial 
errors at SFRs as high as 1 $\msun$ yr$^{-1}$, and biases $\gtrsim 0.5$ dex at the lowest SFRs.
We emphasize that due to the stochastic behavior of blue SFR tracers, one cannot assign a deterministic single value to the 
SFR of an individual galaxy, but we suggest methods by which future analyses that rely on integrated-light 
indicators can properly account for these stochastic effects. 
\end{abstract}


\section{Introduction}\label{sec:intro}

Stellar light is the primary observable in astronomy, and it provides most of our knowledge of the universe and its evolution. While for the nearest stellar populations we can observe individual stars, we are often restricted to measuring the integrated photometric properties of stars, both spatially and spectrally. These integrated properties, when filtered through a model for stellar populations, can then yield estimates of the mass, star formation rate (SFR), star formation history (SFH), initial mass function (IMF), and numerous other properties.

Because the light produced by a star is a function of its mass and age, the stellar population synthesis (SPS) models required to map between observed luminosity and underlying physical properties involve calculating a sum over the mass and ages of all the stars that comprise the population. The most commonly-used approaches for evaluating this sum rely on several assumptions for computational efficiency. Most relevant to this paper, it is common to assume that the IMF and SFH are infinitely well populated (e.g., \texttt{Starburst99}: \citealt{starburst99, sb992}; \texttt{PEGASE}: \citealt{pegase}; \texttt{GALEV}: \citealt{galev}, but see \citealt{anders2013}; \texttt{FSPS}: \citealt{conroy1, conroy2, conroy3}). This approach is convenient because it replaces the sum with a separable double-integral: one first integrates over the IMF at fixed time to calculate the light per unit mass for a stellar population as a function of age, and then integrates this light to mass ratio weighted by the SFH in order to arrive at an estimate of the integrated light produced by stars of all ages.

While this approach is convenient, it can also be dangerous. The potential pitfalls of assuming a fully-sampled IMF when analyzing a simple stellar population (i.e., a group of stars of uniform age) are well-known: if the IMF is not fully-sampled, the highly nonlinear dependence of luminosity on stellar mass causes the manner in which stars discretely fill a population's mass to have large consequences for the luminosity \citep[e.g.,][]{cervino1, cervino3, popescu09a, popescu10a, popescu10b, anders2013}. In this case, there is no longer a deterministic relation between the total mass and age of the population to the total luminosity and color of its integrated light. The implication is that the inverse problem, that of determining the mass or age of a simple stellar population from its photometric properties, no longer has a unique solution. Nor can this non-uniqueness be described as a simple error symmetrically bracketing a central estimate. In a small stellar population, a single high mass star can dramatically increase (and at times dominate) the luminosity of a stellar population. Thus, this very high luminosity for rare realizations skews the mean of the luminosity distribution well away from its median. As a result, mean relations for luminosities that are in the stochastic regime (where a single star can dramatically affect the luminosity) often greatly overpredict the luminosity of a randomly chosen realization. See \citet{cervinoReview} for a recent review of this topic, and a discussion of the implications of these uncertainties.

The hazards of assuming a smooth SFH over timescales of few million years, and thus the accuracy of SPS models that make this assumption, have received significantly less attention (e.g., see the recent review by \citealt{kennicutt12a}; see also \citealt{cen14}). We know from observations of both the Milky Way and nearby galaxies that star formation is a highly clustered process \citep[e.g.,][]{ladalada}, which more closely resembles a series of discrete bursts identifiable with the formation of individual clusters than the continuous creation of new stars at a constant rate. Only when the SFR is sufficiently high do the individual bursts blur together to create an approximately continuous SFH (see figures 3 and 11 of \citealt{slug1}, hereafter Paper I). The questions of how star formation proceeds 
inside stellar clusters populated from a finite IMF sampling, and how clustered star formation affects 
the observed integrated light and the inferred SFHs, motivated us to create the Stochastically Lighting Up Galaxies (\slug) code, presented
in Paper I. This code hierarchically follows clusters drawn from a cluster mass function, each of which is individually populated on a star-by-star basis according to an IMF. Each star evolves following an individual evolutionary track, and contributes light calculated from an individual stellar atmosphere model. As a result of this approach, \slug~produces Monte Carlo realizations of stellar populations rather than simply the mean results, including stochasticity in both the IMF and the SFH. Our initial application of this code 
\citep{miki,slug1}
showed that, for non-simple stellar populations, SFH sampling stochasticity turns out to affect the light output of stellar populations far more than IMF sampling stochasticity. 
Indeed, \citet{mikiletter} (also see \citealt{weiszNew}) show that this effect explains the low H$\alpha$ to FUV ratios seen in dwarf galaxies \citep{lee2009,boselli,meurer}, something that some earlier authors had attributed to variations in the IMF itself.
Since this initial application, \slug~has been used to study these effects in a number of other contexts \citep{siana10a, cook12a, andrews, lyaApp}.

In this paper we extend the application of \slug~to the problem of interpreting SFR indicators (SFIs). These are, by construction, extremely sensitive to the properties of the most massive, shortest lived, brightest stars, and thus are very vulnerable to stochasticity. They are therefore subject to the same ``inverse problem" that affects the determination of mass and age for simple stellar populations: at low SFRs, where IMF and SFH are sparsely sampled, there is no unique mapping between SFRs and SFIs, and thus no unique way to infer a SFR from a SFI in an individual galaxy\footnote{The mean relations are still accurate. On average those SFRs produce that SFI luminosity. However, the interpretations that simply use the mean relation are not appropriate. The broad and highly skewed nature of the PDFs for SFR given a SFI mean
that care must be taken to properly interpret observations.}. Given these limitations, our goal in this paper is to provide the next-best possible solution: a full characterization of the probability distribution function (PDF) of SFR given a particular observed value of SFI. The layout of the paper is as follows: Section \ref{sec:indicate} describes a library of \slug~simulations that we have performed to solve the forward problem of characterizing the distribution of luminosities that result from stochastic sampling of the IMF, including the effects of clustering\footnote{Stellar clustering is the dominant mechanism for SFH sampling stochasticity, thus in several places we use the terms ``SFH stochasticity" and ``effects of clustering" interchangeably.} and a discussion of the dependence on free parameters. Section \ref{sec:miner} describes how we use these \slug~simulations to solve the inverse problem of determining the PDF of SFR given a set of observations, including the higher-dimensional correlations between the true underlying SFR and multiple SFIs. Finally, Section \ref{sec:discussion} discusses the implications of this work, and Section \ref{sec:summary} summarizes our conclusions.

\section{The Distribution of Luminosity at Fixed Star Formation Rate}\label{sec:indicate}

\subsection{\texttt{SLUG} Simulations}

We first consider the problem of determining the distribution of luminosities of SFIs given an input SFR. This allows us to determine, for example, how much scatter is expected for a given stellar population and to characterize the types of errors one might incur if only using the mean properties. We approach this problem via \slugt~simulations, which produce Monte Carlo realizations of photometric properties given a set of user inputs including the input SFH, IMF, the initial cluster mass function (ICMF), the fraction of star formation occurring in clusters, and a  set of stellar evolutionary tracks and atmosphere models. The code also takes parameters describing how clusters disrupt, but these affect only the properties of the cluster population, not the integrated light of a galaxy, and so we will not refer to them further. Unless otherwise noted, all our simulations make use of the default \slug~parameter choices described in Paper I, and summarized in Table \ref{table:stellarprop}. We also refer the readers to \citet{slug1} for a full description of \slug's functionality,
but we provide here a brief summary on how the IMF and ICMF are implemented, which is most relevant to our discussion.

\slug\ treats both the IMF and ICMF as probability distribution functions. When clustered star formation 
is enabled (i.e. $f_{\rm c}>0$), \slug\ draws clusters from the ICMF, at a rate which satisfies 
the imposed SFH. Each cluster is then populated with stars drawn at random from the IMF, until 
the collective mass of the drawn stars equal to the cluster mass. For the last star, \slug\ implements a 
so-called ``stop-nearest'' method \citep{haas2010}, according to which the last drawn star  is included only
if keeping it makes the total effective mass of the cluster closer to the one drawn from the ICMF 
than leaving it out. Following this algorithm, the effective IMF (i.e. the IMF which results from 
repeated draws of an input Kroupa IMF) differs on average from the input IMF due to the constraints imposed 
by the cluster mass distribution. As shown in the previous papers of this series \citep{mikiletter,slug1},
these additional constraints enhance the effect of stochasticity compared to random drawing of the 
IMF alone. 

As discussed in our previous work, this approach should not be confused with other 
formulations which introduce a deterministic relationship between the galaxy SFR and 
its IMF or ICMF. A popular example of one such formulation is the original implementation of the
IGIMF \citep[e.g.][]{igimf0,weid2004,weid2010}, in which a truncation of both the IMF and ICMF is set 
by deterministic equations. As shown in \citet{mikiletter}, this approach differs from \slug\ simulations,
yielding flux distributions that systematically miss the high fluxes associated to the presence of massive
stars. Conversely, these high fluxes are still achievable in \slug\ simulations, although with low
probability. In particular, we found that the deterministic limitation on stellar masses introduced in the IGIMF models actually leads to a dramatic reduction in the luminosity scatter at low SFRs compared to our model of random sampling.  An IGIMF-like model would therefore produce significantly smaller scatters but significantly higher biases than the fiducial results we present below, as one can see from the distributions presented in \citet{mikiletter}. However, we do not explore this topic further, because  \citet{mikiletter} found that the reduced scatter predicted by deterministic IGIMF models is strongly inconsistent with observational constraints \citep[see also][]{andrews,andr14}.

\begin{table}
\caption{\slug~Simulation Parameters.}\label{table:stellarprop}
\begin{center}
\begin{tabular}{lc}
\hline
 & Fiducial\\
\hline
$f_c$ & 1  \\
$t_{\rm sf}$ [Myr] & 500\\
$m_{\rm max}\, [\msun]$& $10^7$  \\
$m_{\rm min}\, [\msun]$& $20$  \\
$[{\rm Fe}/{\rm H}]$ & 0 \\
IMF & Kroupa \\
\hline
\end{tabular}
\end{center}

Here $f_c$ is the clustering fraction, $t_{\rm sf}$ is the duration of star formation, $m_{\rm min}$ and $m_{\rm max}$ are the minimum and maximum of the ICMF, $[{\rm Fe}/{\rm H}]$ is the metallicity used for the stellar evolution and atmosphere models, and IMF is the choice of stellar initial mass function. For a description of how each of these parameters is implemented in \slug, see Paper I.
\end{table}

For the purposes of this paper, we restrict ourselves to very simple input SFHs: those with constant SFR over a time of 500 Myr.\footnote{It is important to note that, as discussed in Paper I, the input SFH does not match the actual realized SFH. 
In fact, due to stochastic sampling of the cluster mass function, the output SFH will differ from the input SFH as it will exhibit a series of bursts on small time scales (see figure 3 in Paper I).
This is because there is no ``constant" SFR. For example, consider a galaxy forming stars at 1 $\msun\pyr$. In one day, 1/365th of a solar mass of gas is not transformed into a star. Constant SFRs (and SFHs in general) can only be considered continuous when averaged over some time interval. In our case, the observations dictate their own averaging window and we investigate how well the continuous model matches reality.}  Our choice of time period is long enough that we avoid any transient initial phases of the buildup of the stellar population. The primary output of each \slug~simulation is a realization of the PDF of luminosities given a SFR and other ancillary variables,
\begin{equation}
p(\vl \mid \log \sfr, \phi),
\end{equation}
 where  $\vl$ is a vector of log luminosities in various photometric bands and $\phi$ denotes parameters that define the model,
as listed for example in Table \ref{table:stellarprop}.
For simplicity, in the analysis that follows we will omit $\phi$ except where relevant. 
 
While \slug \, is capable of producing photometry in many bands, and the next release of the code will support full spectra, here we focus on the three most common indicators of the SFR: the FUV luminosity $L_{\rm FUV}$, the bolometric luminosity $L_{\rm bol}$, and the H$\alpha$ luminosity $L_{\rm H\alpha}$. The last of these is a recombination line produced when the ionizing radiation of the stars interacts with the ISM, and \slug~does not report this directly. Instead, it reports the rate of hydrogen-ionizing photon emission $\qho$, which we convert to H$\alpha$ luminosity via
\begin{eqnarray}
L_{\mha}&=&(1-f_{\rm esc})(1-f_{\rm dust})\textrm{Q(H}^0) \alpha_{\mha}^{\textrm{eff}} h\nu_{\mha} \nonumber \\
&\approx& 1.37 \times 10^{-12} (1-f_{\rm esc})(1-f_{\rm dust})\textrm{Q(H}^0)\mbox{ erg},
\end{eqnarray}
where $f_{\rm esc}$ and $f_{\rm dust}$ are the fractions of ionizing photons that escape from the galaxy and that are absorbed by dust grains rather than hydrogen atoms, respectively, $\alpha_{\mha}^{\rm eff}$ is the recombination rate coefficient for recombination routes that lead to emission of an H$\alpha$ photon, and $h \nu_{\mha} = 1.89$ eV is the energy of an H$\alpha$ photon.  For the purposes of our analysis, we assume $f_{\rm esc} = 0$, noting that non-zero values for this poorly-constrained quantity would simply amount to applying a constant shift to our results, provided that $f_{\rm esc}$ does not vary substantially with galaxy properties \citep[see a discussion in e.g.][]{boselli}. Similarly, although we focus on H$\alpha$, the results will be identical up to a constant shift for any other hydrogen recombination line, or any other source of emission (e.g., free-free emission) that is directly proportional to the ionizing luminosity. We leave for future work the discussion of other SFIs that have more complex, non-linear relationships with the ionizing photon production rate (e.g., [O~\textsc{ii}] 372.7 nm, [Ne~\textsc{ii}] 12.8 $\mu$m, [Ne~\textsc{iii}] 15.6 $\mu$m -- \citealt{kennicutt12a}). 
In this work, for all indicators, we report intrinsic luminosities in the absence of dust (i.e. 
assuming $f_{\rm dust}=0$ in eq. 2). Therefore, results from our study need to be compared to dust-corrected fluxes, 
after accounting for dust absorption of the selected indicator and, for H$\alpha$, also for the intrinsic absorption of the
ionizing photons within HII regions. As shown in e.g. \citet{boselli}, a particular choice of the many available dust extinction 
laws \citep[e.g.][]{bua05,meur1999,calzetti2001} may imprint non-negligible systematics on the resulting UV and H$\alpha$ 
flux distribution, requiring particular care in the derivation of unobscured fluxes before comparing to our theoretical models.

\begin{figure}
\includegraphics[scale=0.75]{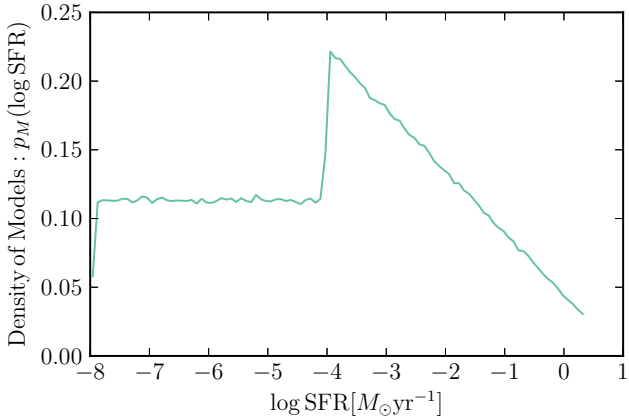}
\caption{Distribution of input SFRs for \slug~simulations. The sudden discontinuity at $10^{-4}$ occurs simply because we
compute a different number of models above and below $10^{-4}~\rm M_\odot~yr^{-1}$. As 
discussed in the text, this choice enables a more efficient simulation strategy, but it does not affect our results.}
\label{fig:modelsfr}
\end{figure}

In order to characterize the PDFs of our chosen SFIs, we run approximately $1.8\times 10^6$ \slug\ models. Of these models, we 
run $9.83\times 10^5$ at input SFRs with a distribution of $\log\sfr$ that has a linear form with a slope of -1 over a range 
in $\log{\rm SFR}$ from $-4$ to $0.3$, where SFRs here are measured in $M_\odot$ yr$^{-1}$. The remaining $0.8\times 10^6$ models 
are uniformly distributed in $\log{\rm SFR}$ over a range from $-8$ to $-4$. The distribution of the model SFRs 
$p_{M}(\log\sfr)$ is shown in Figure \ref{fig:modelsfr}. Our choice of distribution is motivated by the practical requirement 
that we need more simulations to adequately sample the PDFs at lower SFRs because the scatter is larger. As we will show in 
section \ref{ssec:derivation}, our results do not depend on the assumed distribution of models, $p_M(\log \sfr)$.

\subsection{Simulation Results}\label{sec:depend}

For convenience, we report the result of our simulations in SFR space, meaning that we report luminosities as the SFRs one would infer using the approximation of perfect IMF and SFH sampling, which we refer to as the ``point mass approximation"\footnote{This is standard statistical terminology. It arises from the fact that one treats the posterior PDF as having all of its mass located at a single point.}. For our fiducial IMF, stellar evolution tracks and atmosphere models, the conversions between these and the luminosities reported by \slug~are
 \begin{eqnarray}
 \label{eq:sfrho}
\sfr_{Q(\rm H^0)} &= &  7.638 \times 10^{-54} (\msun {\rm yr}^{-1}\ {\rm s}) Q({\rm H}^0)\\
\sfr_{\rm FUV} &=& 9.641\times10^{-29} (\msun {\rm yr}^{-1} \ {\rm erg}^{-1} \ {\rm s}\ {\rm Hz}) L_{\rm FUV}\\
\sfr_{\rm bol} &= & 2.661\times10^{-44} (\msun {\rm yr}^{-1} \ {\rm erg}^{-1} \ {\rm s} ) L_{\rm bol}.
\label{eq:sfrbol}
  \end{eqnarray}
This approach allows us to report the results using the different SFIs on a common scale, making them easier to compare. It also allows us to separate the effects of stochastic sampling from the dependence of the results on the choice of stellar evolution and atmosphere models as these, to good approximation, simply cause changes in the conversion constants in equations (\ref{eq:sfrho}) -- (\ref{eq:sfrbol}).

\begin{figure*}
\begin{tabular}{ccc}
\includegraphics[scale=0.7]{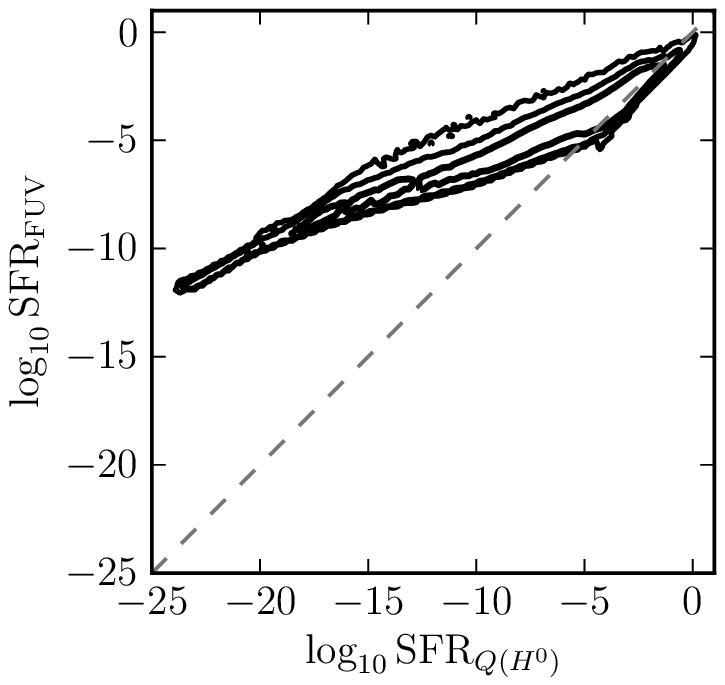}&
\includegraphics[scale=0.7]{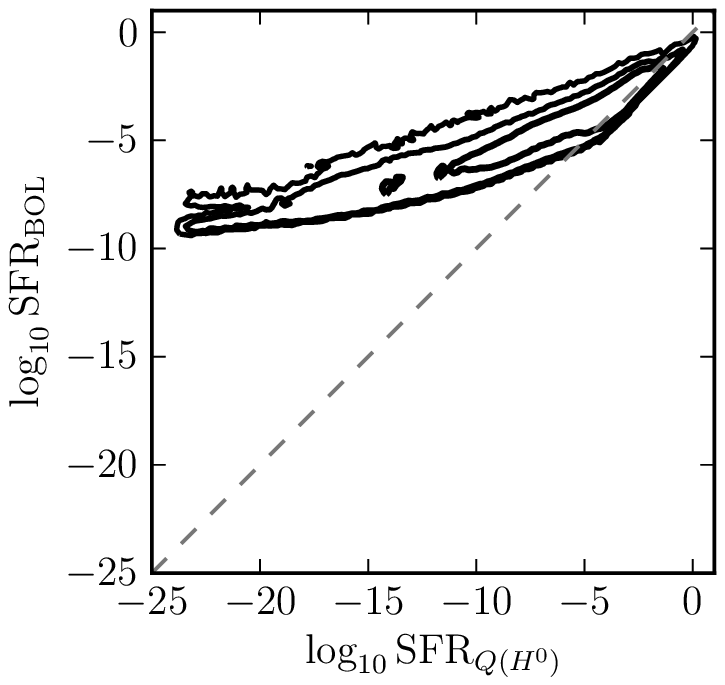}&
\includegraphics[scale=0.7]{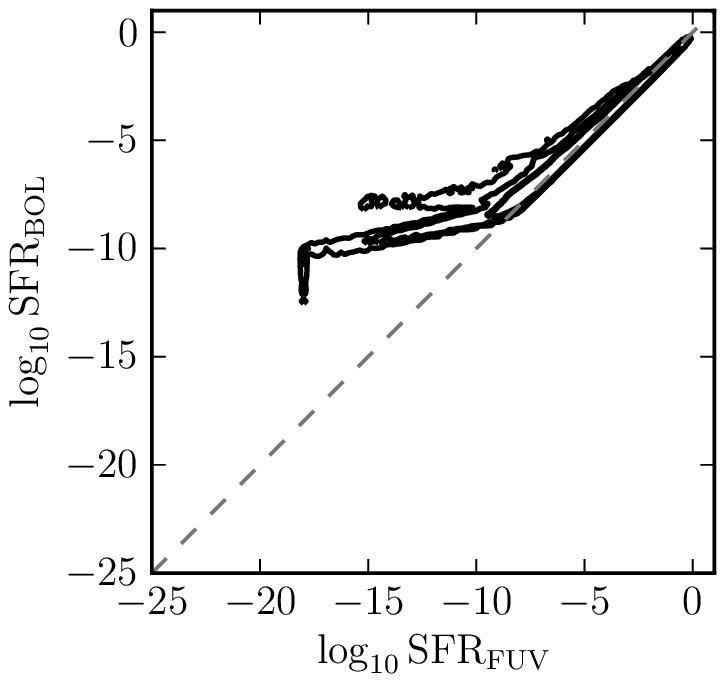}
\end{tabular}
\caption{Contours representing $10^{[1,2,3]}$ models for the three different SFIs converted to SFRs using the point-mass approximations. Without stochastic effects, the galaxies would be forced to lie exactly on the dashed line.}
\label{fig:pdfs}
\end{figure*}

\begin{figure*}
\begin{tabular}{cc}
\includegraphics[scale=0.9]{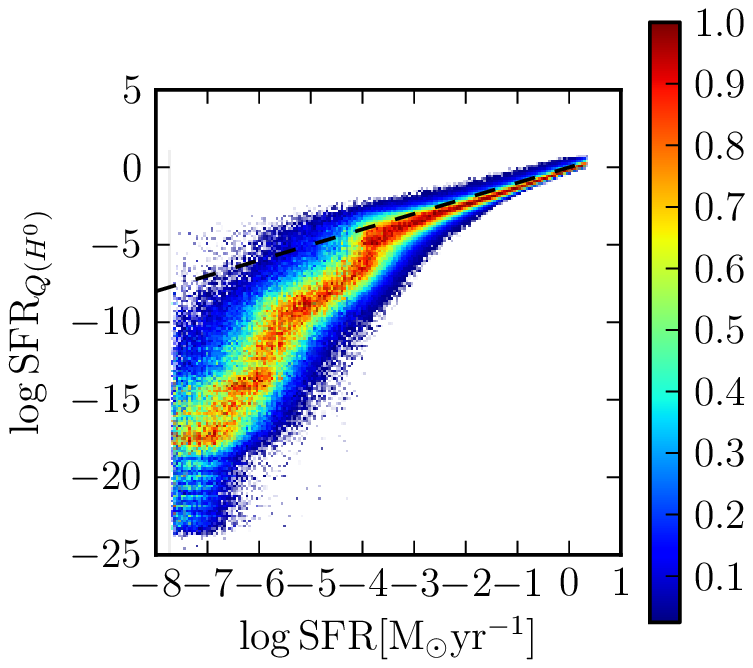}&
\includegraphics[scale=0.9]{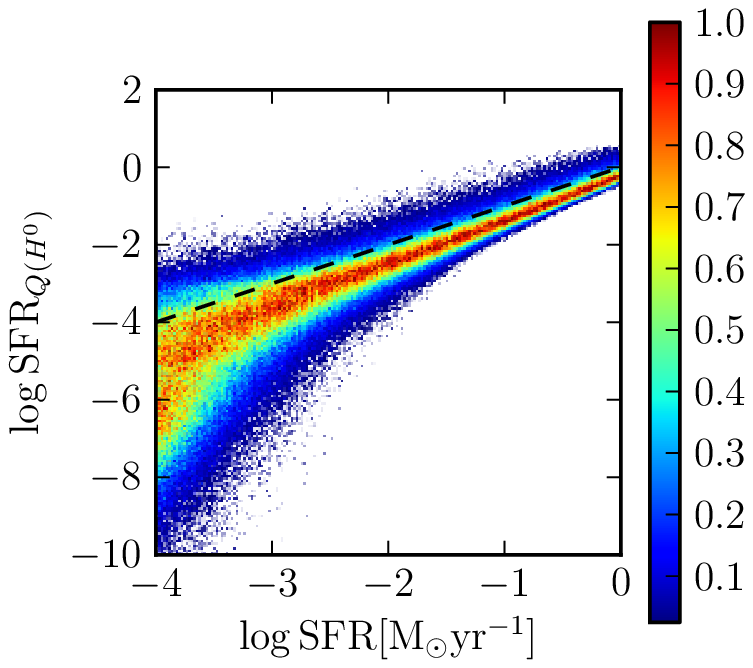}\\
\includegraphics[scale=0.9]{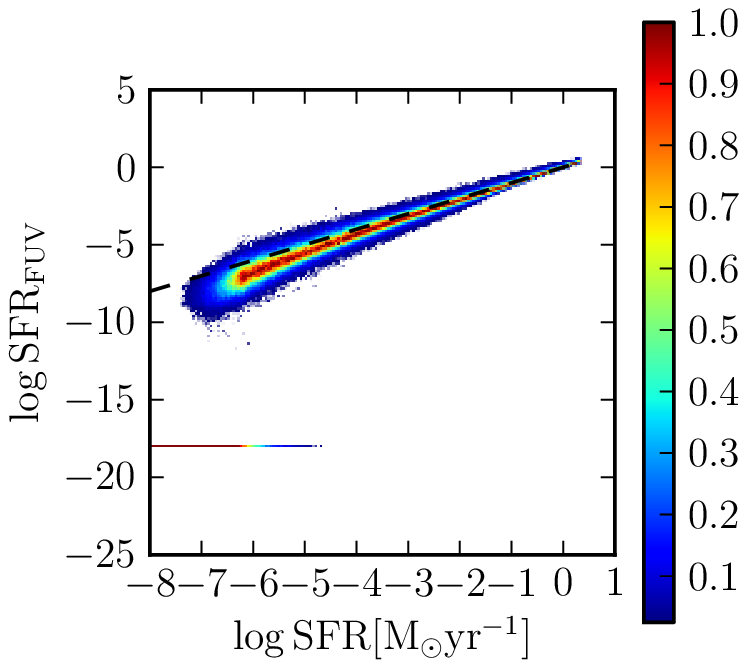}&
\includegraphics[scale=0.9]{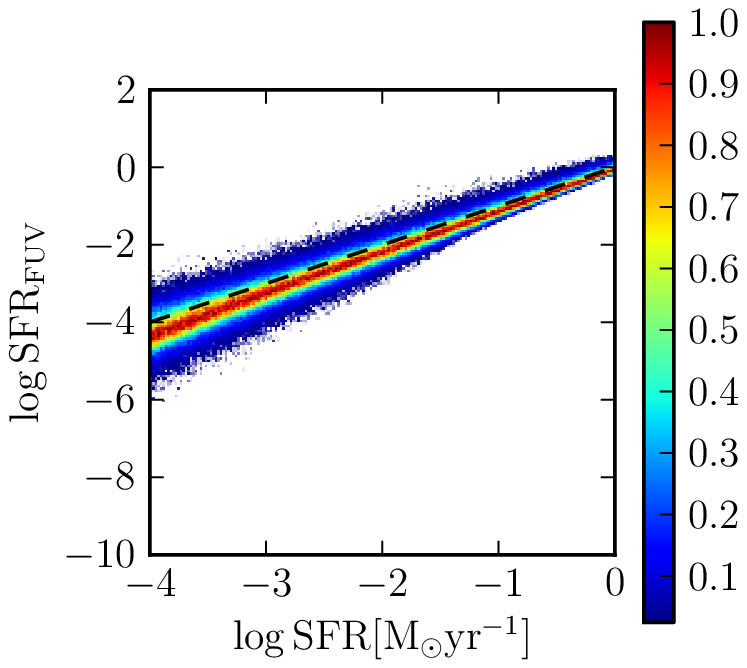}\\
\includegraphics[scale=0.9]{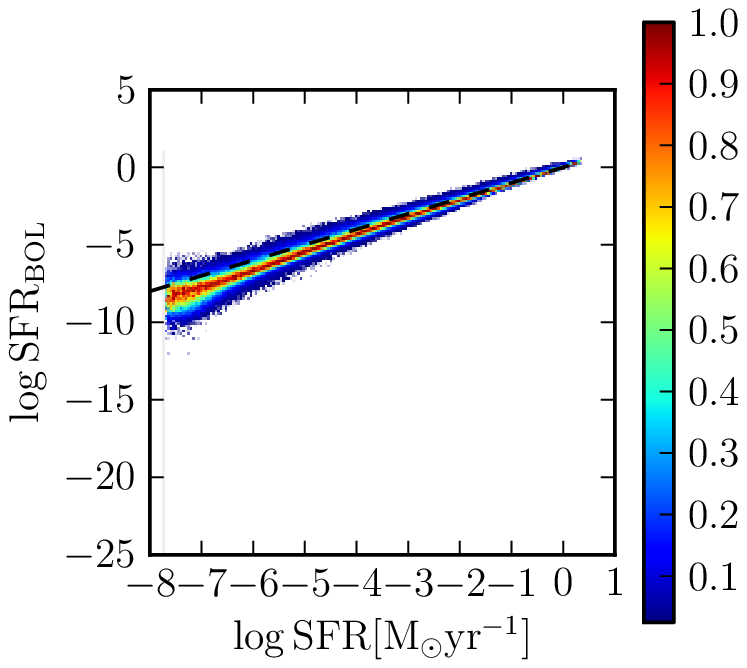}&
\includegraphics[scale=0.9]{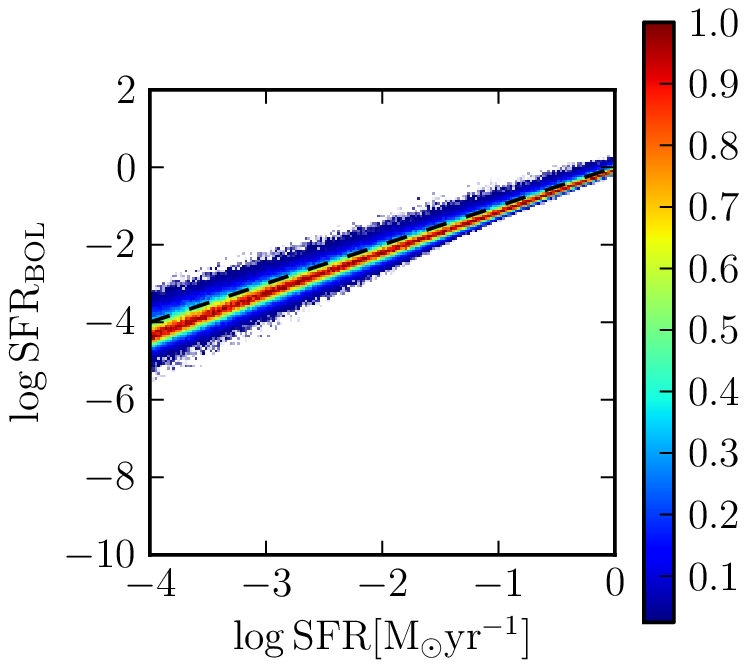}\\
\end{tabular}
\caption{(\emph{left}) PDFs of the SFIs vs.~intrinsic SFR arising just from stochastic effects (presented as fraction of the maximum value in each intrinsic SFR bin). The dashed line represents the point-mass approximation. The hard cutoff at $\log\sfr=-8+\log 2 [\msun \pyr]$ is the smallest SFR that can produce 
any clusters with a mass of 20 $M_\odot$, the minimum cluster mass we allow. 
The horizontal stripe for $\sfrfuv$ at -18 corresponds to the lower limit of FUV luminosity given by the \slug \, models. (\emph{right}) Zoomed in version of plots in left column.}
\label{fig:bigiel}
\end{figure*}

Each \slug~model may be thought of as a point in a four-dimensional parameter space defined by these three luminosities and their corresponding intrinsic SFR. In Figure \ref{fig:pdfs}, we show the raw distribution of our models in three orthogonal projections of 
this parameter space. One can see that at progressively lower fluxes, which correspond to progressively lower intrinsic
SFRs, the SFRs inferred applying the point-mass approximation deviate from the 1:1 relation, emphasizing 
the need for accurate modeling in the conversion between SFIs and SFRs in this regime. One can also see that the 
effect is more pronounced for indicators which are most sensitive to massive short-lived stars, particularly $Q({\rm H}^0)$.
Figure \ref{fig:bigiel} presents the distributions related to their intrinsic SFRs.
We can immediately see that there is significant mass of models well away from the line predicted by the point-mass approximation, confirming the necessity of the stochastic treatment and our assertions that full PDFs should be used in place of simple mean relations. We also see that, as expected, the deviation from the line is largest for $\sfr_{Q({\rm H}^0)}$, and smaller for the other two dimensions. This was discussed in \cite{mikiletter}, as tracers that are sensitive to stars with lifetimes shorter than a few Myr are most sensitive to the flickering in the SFH, while SFI that depend on longer lived stars average over longer time scales and are thus more stable in recovering the mean SFH. In passing, we note that at progressively lower SFRs, an increasingly higher number of models do not contain 
massive enough stars that produce UV fluxes in \slug\ models. These realizations are set to a floor of 
$\log \sfrfuv = -18$ and are shown in Figure \ref{fig:bigiel} to preserve the correct number density of models 
in each SFR bin.

\begin{figure}
\begin{tabular}{c}
\includegraphics[scale=0.7]{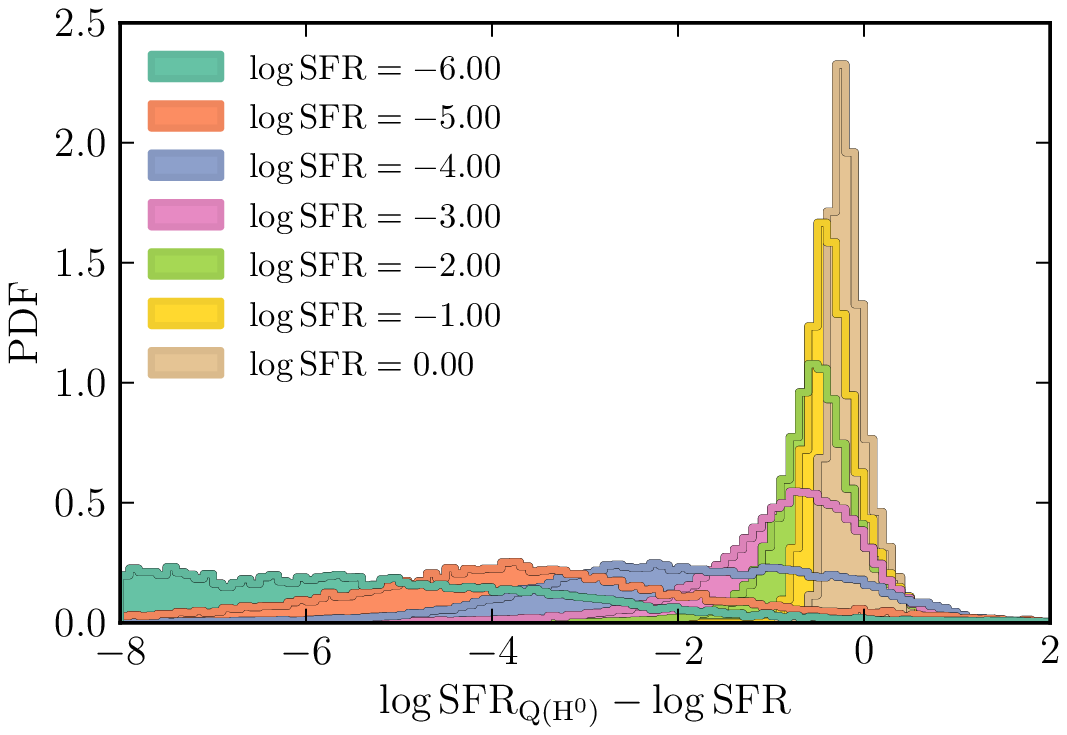}\\
\includegraphics[scale=0.7]{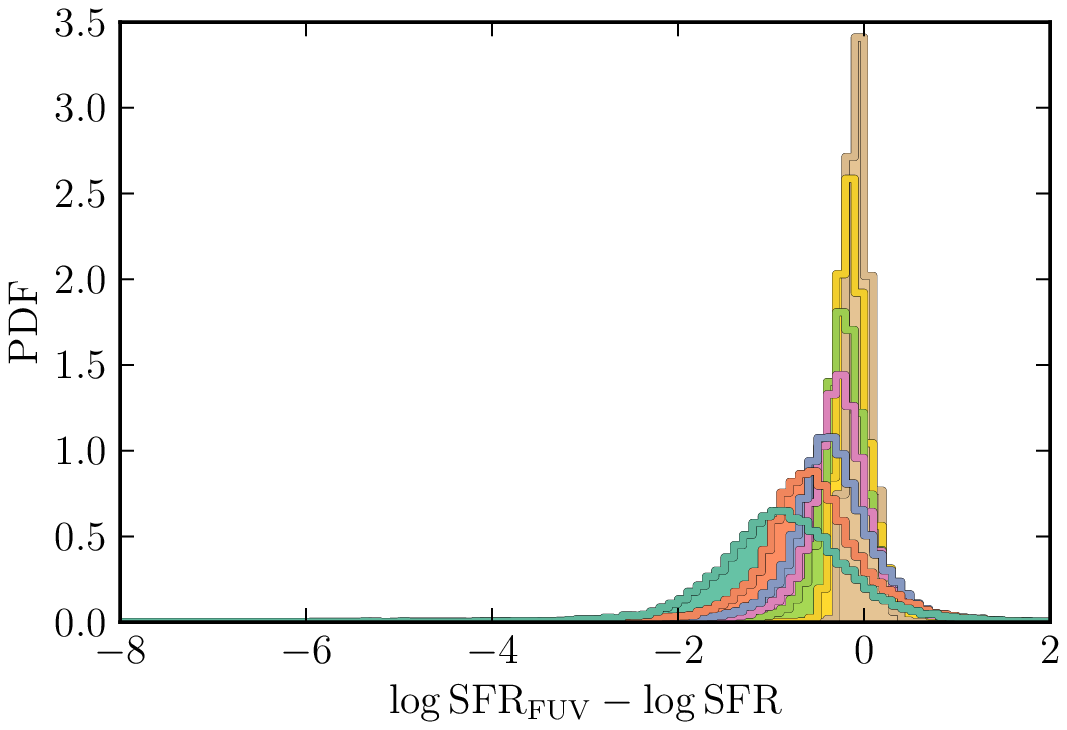}\\
\includegraphics[scale=0.7]{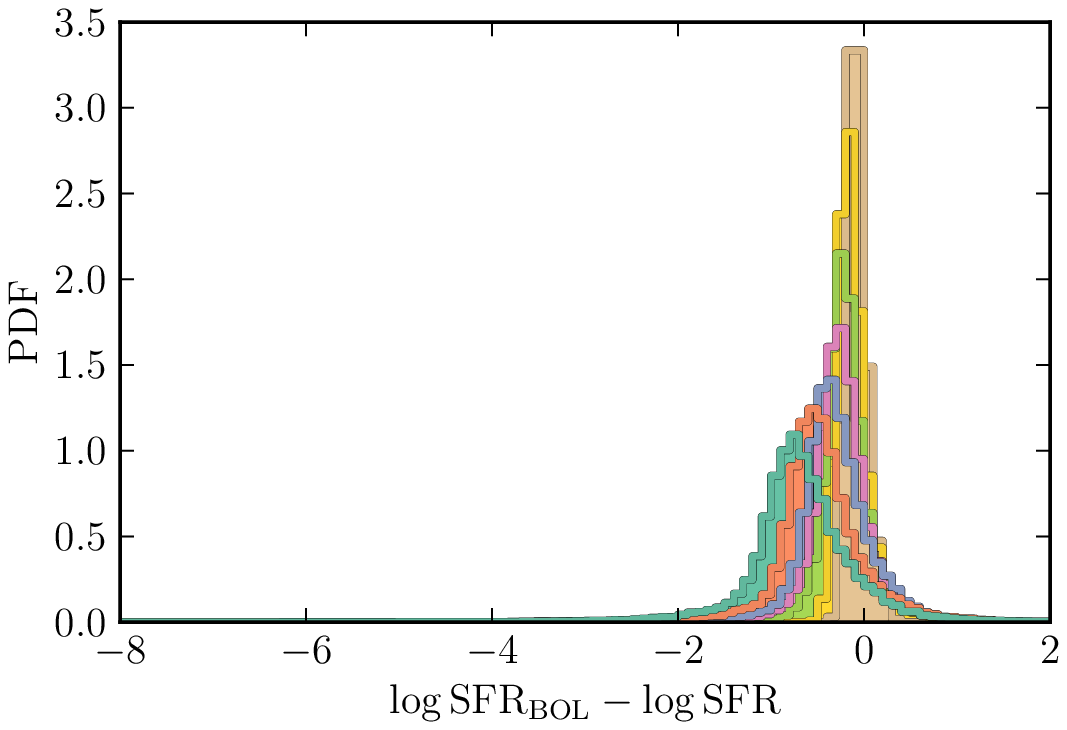}
\end{tabular}
\caption{PDFs for individual components of $\vl$ normalized by the point-mass approximation for ease of comparison. Models
are grouped by SFR into bins $0.25$ dex wide, and are color-coded by input SFR as indicated in the legend.}
\label{fig:sfi_pdfs}
\end{figure}

While a clear picture of the ensemble of all the models is presented in Figure \ref{fig:bigiel} (which is critically useful in our subsequent analysis -- see Section \ref{sec:miner}), explorations of the level of scatter can perhaps be better addressed by Figure \ref{fig:sfi_pdfs}, which shows the marginal distributions of $p(\vl\mid\log\sfr)$. 
To emphasize the shape of the distribution over the actual values that are related to adopted point mass calibrations,
we plot the distribution of the offsets between these inferred SFRs and the true SFR that was used in each simulation. It is again clear that $\qho$ has the largest scatter\footnote{
It is important to caution that, while large scatter is a real limitation of SFIs based on ionizing luminosity, it would be incorrect to conclude from this that alternates such as FUV or bolometric luminosity are always preferable. If the true SFR is stable on the $\sim 10-100$ Myr timescales to which these tracers are sensitive, as is the case in our models, then, all other factors equal, they are preferable. However, in a galaxy where the intrinsic SFR might be variable on shorter timescales (e.g., in a merging or interacting galaxy), the longer averaging interval of FUV or bolometric luminosity becomes a disadvantage, as it produces too coarse an estimate of the true recent SFR.},
 in extreme cases producing estimates that differ from the true SFR by as much as eight orders of magnitude!  Furthermore, these distributions are clearly not Gaussians centered on the true SFR. Instead, they are highly asymmetric. Finally, it is clear that as the SFR increases, the PDF gets narrower. This is the result of being better sampled and the laws of statistics of large numbers.

\section{The Distribution of Star Formation Rate at Fixed Luminosity}\label{sec:miner}

\subsection{Derivation}\label{ssec:derivation}

Thus far we have shown how one can estimate the probability distribution of log luminosities $\vl$ given an intrinsic SFR, $p(\vl\mid\log\sfr)$. However, we want to invert the problem and find the full distribution of SFRs given $\vl$, i.e., $p(\log \sfr\mid \vl)$. We perform this inversion with a technique known as implied conditional regression. The idea behind this technique is simple. We start  with the following
decomposition:
\begin{equation}
 p(\log\sfr\mid \vl) = \frac{ p(\log\sfr,\vl)}{p(\vl)}.
 \label{eqt:decomp}
\end{equation}
Each \slug~model has a known SFR and produces an output $\vl$, and thus represents a sample point in the multidimensional parameter space $(\log\sfr, \vl)$; we denote such a point as a vector $\mathbf{z}$, where the first component is $\log\sfr$, and the three log luminosities that comprise $\vl$ form the second through fourth components. This definition can obviously be generalized to an arbitrary number of components in $\vl$. In this space, we define the distance between two points $\mathbf{z}_1$ and $\mathbf{z}_2$ by the usual Cartesian metric,
\begin{equation}
\begin{split}
|\mathbf{z_1} - \mathbf{z}_2| = [(\log\sfr_1-\log\sfr_2)^2 \\
+(\log\sfr_{Q({\rm H}^0), 1} - \log\sfr_{Q({\rm H}^0), 2})^2 + \cdots]^{1/2}.
\end{split}
\end{equation}

The first task in computing $p(\log\sfr\mid \vl)$ is to use these sample points to estimate the underlying multidimensional probability distribution $p(\log\sfr,\vl)$ and its projection along the $\log\sfr$ direction
\begin{equation}
\label{eqt:pdfprojection}
p(\vl) = \int p(\log\sfr,\vl) \, {\rm d}\log\sfr.
\end{equation}
To do this, we use a kernel density estimation technique which constructs the PDF as a sum of kernels centered on each multidimensional simulation point. Explicitly, we approximate the value of the PDF at a position $\mathbf{z} = (\log\sfr,\vl)$ by
\begin{equation}
\label{eqt:pdfestimate}
p(\log\sfr,\vl) = A \sum_i K( |\mathbf{z}_i - \mathbf{z}|;  h),
\end{equation}
where $\mathbf{z}_i$ is the position of the $i$th sample point, $A$ is a normalization constant, and $K$ is the kernel function, which has the bandwidth parameter $h$. For its compactness, we choose to use an Epanechnikov kernel, which is of the form
\begin{equation}
K(z; h) \propto \left\{
\begin{array}{ll}
1-z^2/h^2, & z < h \\
0, & z\geq h
\end{array}
\right..
\end{equation} 
The parameter $h$ must be chosen to balance the competing demands of smoothness, favoring larger $h$, and fidelity, favoring smaller $h$. We choose to set this parameter equal to 0.1 dex because exploration of histograms at various bin sizes indicates that there is little structure below this scale. We are thus washing out any features of this PDF on scales below 0.1 dex in any dimension. The result of this procedure is an estimate of the multidimensional probability density $p(\log\sfr,\vl)$ describing our raw \slug~data, and, by plugging into equation (\ref{eqt:decomp}), an estimate of $p(\log\sfr\mid \vl)$.

The second step in computing $p(\log \sfr\mid \vl)$ is to the apply a proper weighting of the prior probability distribution of 
SFRs. Simply applying equation (\ref{eqt:decomp}) using our computed $p(\log\sfr,\vl)$ amounts to adopting a prior probability distribution of the logarithmic SFR that follows the distribution of our \slug~simulations, shown in Figure \ref{fig:modelsfr}. This is clearly not an ideal choice, as this distribution was chosen to ensure good sampling of the PDF, rather than to reflect a realistic prior distribution. Fortunately, it is trivial to rescale the results to an arbitrary prior probability distribution using Bayes's theorem,
\begin{equation}
p(\log\sfr\mid \vl) = \frac{p(\vl\mid\log\sfr)p(\log\sfr)}{p(\vl)},
\end{equation}
where $p(\log\sfr)$ is the prior probability distribution for the SFR.

Our input grid of models has a distribution of $\log\sfr$ given by $p(\log\sfr) = p_M(\log\sfr)$, where $p_M(\log\sfr)$ is the distribution shown in Figure \ref{fig:modelsfr}. Bayes's theorem tells us that we can use the results from one prior distribution $p_1(\log\sfr)$ to find the results for a different prior distribution $p_2(\log\sfr)$ by multiplying $p(\log\sfr\mid\vl)$ by $p_2(\log\sfr)/p_1(\log\sfr)$.\footnote{This operation requires calculation of a new normalization constant, which is simple to compute in the case of the one-dimensional SFR.} For the case of transforming our \slug\, simulations to a desired $p_2(\log\sfr)$, we set $p_1(\log\sfr)=p_M(\log\sfr)$.
This is equivalent to assigning a different relative weighting to each of the models in the library such that the effective $p(\log\sfr)$ matches whatever form is desired. 

Obviously, the relationship between the SFI and the intrinsic SFR will depend on the choice of the prior, which 
should be made according to the problem at hand. For the purposes of this analysis, we present two different examples, 
which allow us to compare results obtained adopting different priors. 
The first prior we adopt is the most natural choice that can be made when analyzing a large sample of 
galaxies from a volume-limited survey. The SFR distribution of galaxies 
in the local universe is observed to follow a Schechter function, with slope $-1.51$ and characteristic SFR of 9.2 $M_\odot$ yr$^{-1}$
according to the determination of \citet{bothwell}. It is therefore natural to assume that the prior distribution of SFR 
follows a similar Schechter function. A caveat to this assumption is that the observational determination 
of the Schechter function parameters was made ignoring the effects of stochasticity. This is unlikely to affect the characteristic SFR, 
since this is high enough that stochastic effects probably do not dominate the error budget for the FUV plus IR SFR indicators used in 
these observational studies (c.f.~Figure \ref{fig:sfi_pdfs}). On the other hand, the slope at low SFRs may be more problematic, 
a topic to which we return below.

To highlight the importance of priors in the final result, we also consider a flat distribution of $\log \sfr$. Flat priors are often assumed to be a ``robust'' or ``agnostic'' choice, but in reality they are neither. They represent a specific choice that may be appropriate or inappropriate depending on the sample being analyzed. Applied to the hypothetical problem of establishing 
the SFR distribution in a volume-limited sample as mentioned above, a flat prior would represent a rather 
poor choice because it neglects the fact that lower values of $\log \sfr$ are more common than larger ones in a volume-limited sample. This in turn exacerbates the problems of bias and scatter that we discuss below. Conversely, a flat prior would be a more suitable choice when dealing with samples selected to be representative of a full range of properties, e.g. galaxy masses, with equal number of objects per selection bin. 
In our work, which does not deal with a specific problem, the flat prior does offer an interesting second choice to 
highlight the sensitivity of results to the applied prior. It also offers the benefit 
that it is perhaps easier to visualize how our results would scale when changing prior,
since the term $p_1(\log\sfr)$ is in this case a constant.

Once a prior has been chosen, we are at last in a position to derive the final PDF of $\log\sfr$ given a set of observations. We can think of a given set of observational data as describing a PDF $p(\vl \mid \mbox{data})$ of luminosities in one or more bands; the simplest case would be an observation of a single tracer which produces a central value of log luminosity with a Gaussian error distribution, in which case $p(\vl\mid\mbox{data})$ is a Gaussian in one dimension (corresponding to the SFI measured) and is flat in the other dimensions (corresponding to SFIs that were not measured). Given the observations, and a choice of prior distribution $p(\log \sfr)$ for the SFR, the final posterior distribution for the SFR is given by applying equation (\ref{eqt:decomp}), rescaling by the chosen prior, and then integrating over the luminosity distribution implied by the data. The result is
\begin{equation}
\label{eqt:sfrpdffinal}
p(\log\sfr\mid {\rm data}) = \int \frac{p(\log\sfr,\vl)}{p(\vl)} \frac{p(\log\sfr)}{p_M(\log\sfr)} p(\vl\mid\mbox{data})\,{\rm d}\vl,
\end{equation}
where $p(\log\sfr,\vl)$ is given by equation (\ref{eqt:pdfestimate}), $p(\vl)$ is given by equation (\ref{eqt:pdfprojection}), and $p_M(\log\sfr)$ is the PDF of SFRs in our \slug~simulations.

\subsection{Results}

\begin{figure}
\begin{tabular}{c}
\includegraphics[scale=0.7]{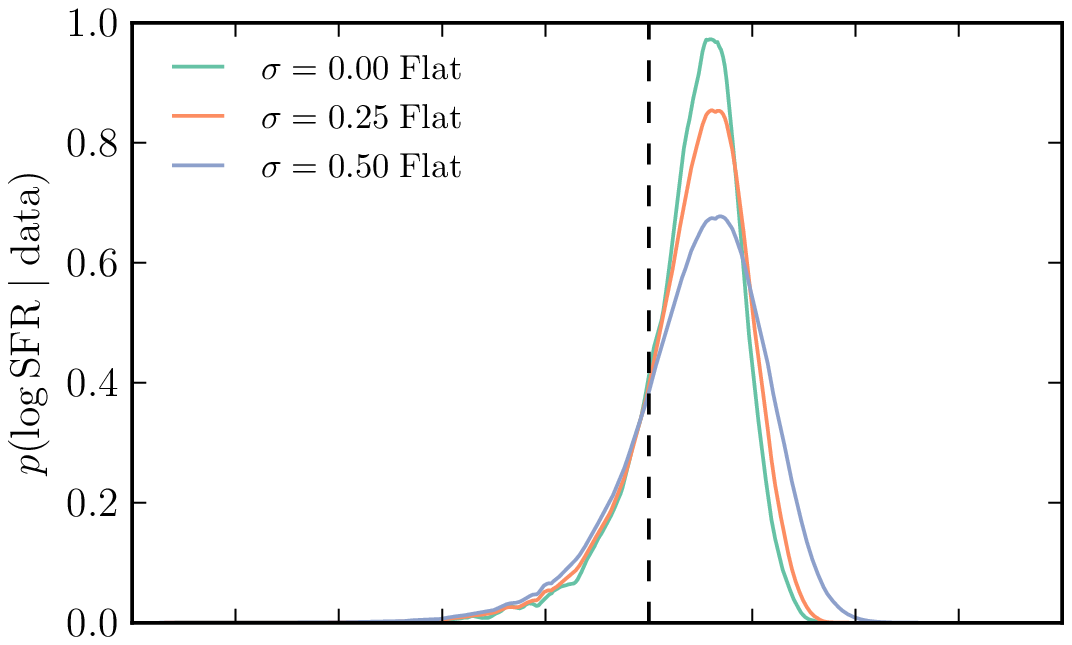}\\
\includegraphics[scale=0.7]{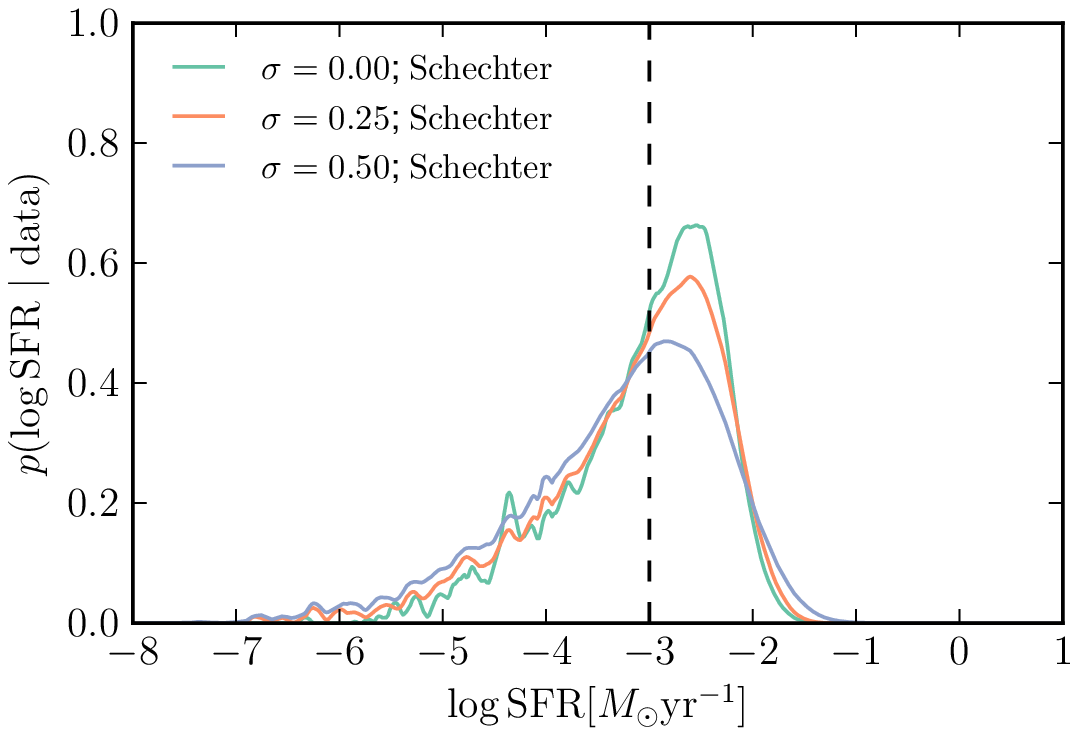}
\end{tabular}
\caption{Posterior distributions for SFR given an observed H$\alpha$ luminosity corresponding to a SFR centered at $\sfrqho=-3$. The observed log luminosity is taken to have a Gaussian-distributed uncertainty whose width $\sigma$ (measured in dex) corresponds to the values shown in the legend; $\sigma=0$ corresponds to a $\delta$ function distribution. The top panel shows results using a flat prior, and the bottom panel shows the results using a Schechter function prior (see Section \ref{ssec:derivation}). The curves get noisier at lower SFRs due to the smaller number of models and the more dispersed nature of the PDFs.} \label{fig:sfrpdf}
\end{figure}

To understand the results for the estimates of $p(\log\sfr\mid\vl)$, we begin by examining an example corresponding to the simplest case of a measurement for a single tracer. Consider an observation of H$\alpha$ luminosity corresponding to $\log\sfrqho=-3$ with a Gaussian error bar of width $\sigma$. In Figure \ref{fig:sfrpdf}, we show the posterior PDF for the SFR given this measurement of H$\alpha$ using both flat and Schechter function priors. If we had to assume point-mass conversion, we would infer $\log\sfr=-3$ for the galaxy SFR (the black dashed line).  However, given the skewness in the flux distribution, the peak and mean of the true PDF\footnote{Note that, as is always the case, this PDF is only true in so far as the prior is the correct prior to use and that other assumptions made are accurate as well regarding the IMF, stellar tracks and atmospheres, etc.} are significantly offset and neither corresponds to the point-mass estimate. We will characterize the difference between the point-mass estimate and the mean of the true PDF as the ``bias". Note that this bias is \emph{not} meant as a simple offset that one can blindly apply to the observational determination to get a ``better" answer that fixes the stochastic issues. In practice, stochasticity fundamentally breaks the deterministic relationship between luminosity and SFR and thus the full PDF should be used whenever possible (or at least the first four moments of the distribution).

We can also see from Figure \ref{fig:sfrpdf} that the posterior PDF of SFR has significant width. Thus even a perfect measurement of the luminosity, corresponding to $\sigma=0$ in the Figure, retains a systematic uncertainty in the SFR with a standard deviation of approximately $0.5$ dex and a significant negative tail. Indeed, in the example shown, this stochastic uncertainty dominates the error budget, as is clear from the fact that the PDFs for observational errors of $\sigma = 0$, $0.25$ dex, and $0.5$ dex are only marginally different. Finally, we can see that the choice of prior does affect the results, but not significantly in this case\footnote{Given that the posteriors are so broad, this is the result of the fact that the priors are similar. Choosing a linearly flat $p(\sfr)\propto 1$ prior would produce significantly different results with a much higher weighting of higher SFRs.}.

Given the results shown in Figure \ref{fig:sfrpdf}, it is obviously of interest to know how the bias and uncertainty depend on the observed value of a particular SFI. We formally define these quantities as follows. Consider an observation of a particular SFI $I$ which returns an estimated logarithmic SFR $\log\sfr_I$ using the point-mass estimate (i.e., using equations \ref{eq:sfrho} -- \ref{eq:sfrbol}), with a Gaussian error distribution $\sigma$ on $\log\sfr_I$. The posterior probability distribution for the true SFR $p(\log\sfr\mid \log\sfr_I\pm \sigma)$ is then given by equation (\ref{eqt:sfrpdffinal}), treating the observed luminosity distribution $p(\vl\mid\mbox{data})$ as a Gaussian of width $\sigma$ centered at $\log\sfr_I$. The corresponding mean estimate of $\log\sfr$ is
\begin{equation}
\begin{split}
\overline{\log\sfr} = \int p(\log\sfr\mid \log\sfr_I\pm\sigma) \log\sfr \, {\rm d}\log\sfr.
\end{split}
\end{equation}
We define the bias $b$ and scatter $s$, respectively, as
\begin{eqnarray}
b(\log\sfr_I) \equiv &  \overline{\log\sfr} - \log\sfr_I \label{eqt:bias}\\
s(\log\sfr_I)^2 \equiv & \int p(\log\sfr\mid \log\sfr_I) \nonumber\\
                 &\left(\log\sfr - \overline{\log\sfr}\right)^2 \, {\rm d}\log\sfr \label{eqt:scatter},
\end{eqnarray}
i.e., for a given observation of a single tracer, we define the bias as the difference between the mean value of $\log\sfr$ computed from the full PDF and the point-mass estimate, and the scatter as the second moment of the PDF of $\log\sfr$. Due to the nature of the distributions, normally the bias is positive.

\begin{figure*}
\begin{tabular}{cc}
\includegraphics[scale=0.75]{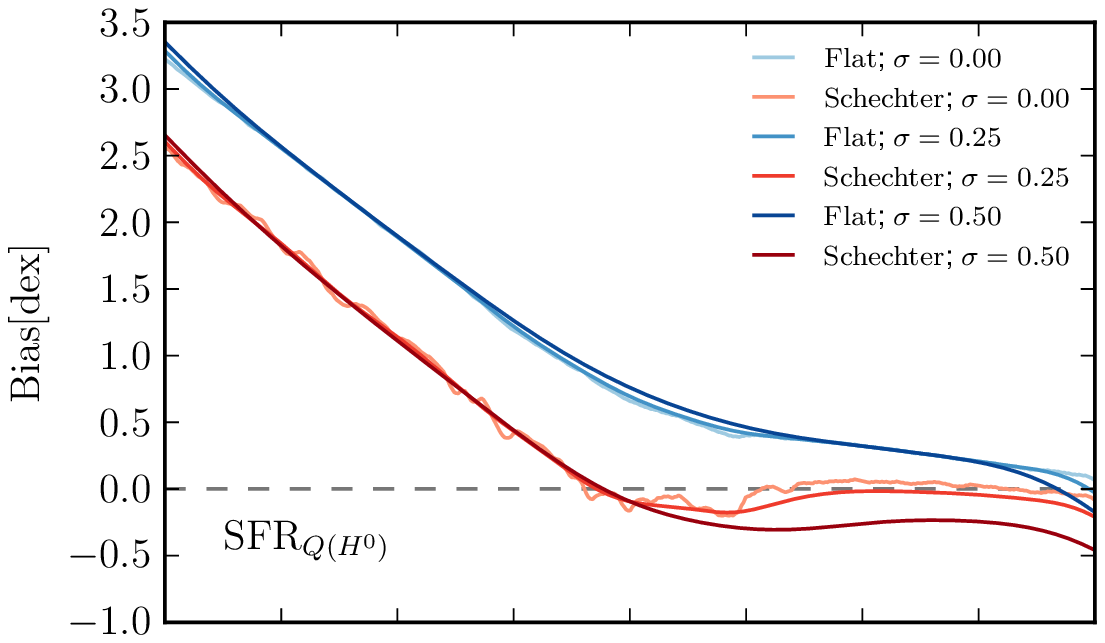} & 
\includegraphics[scale=0.75]{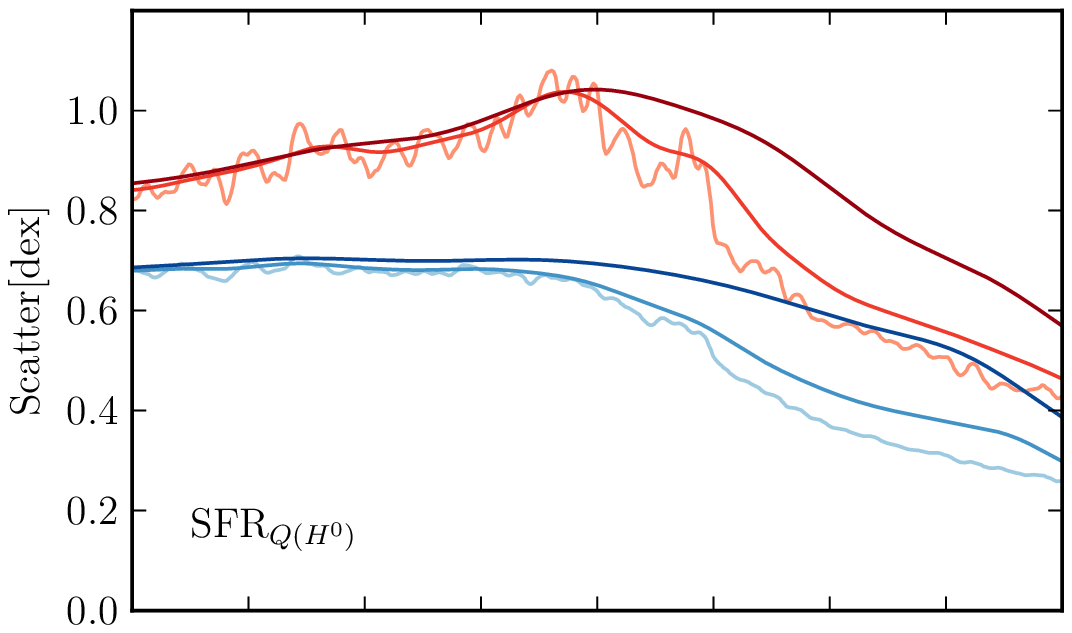} \\
\includegraphics[scale=0.75]{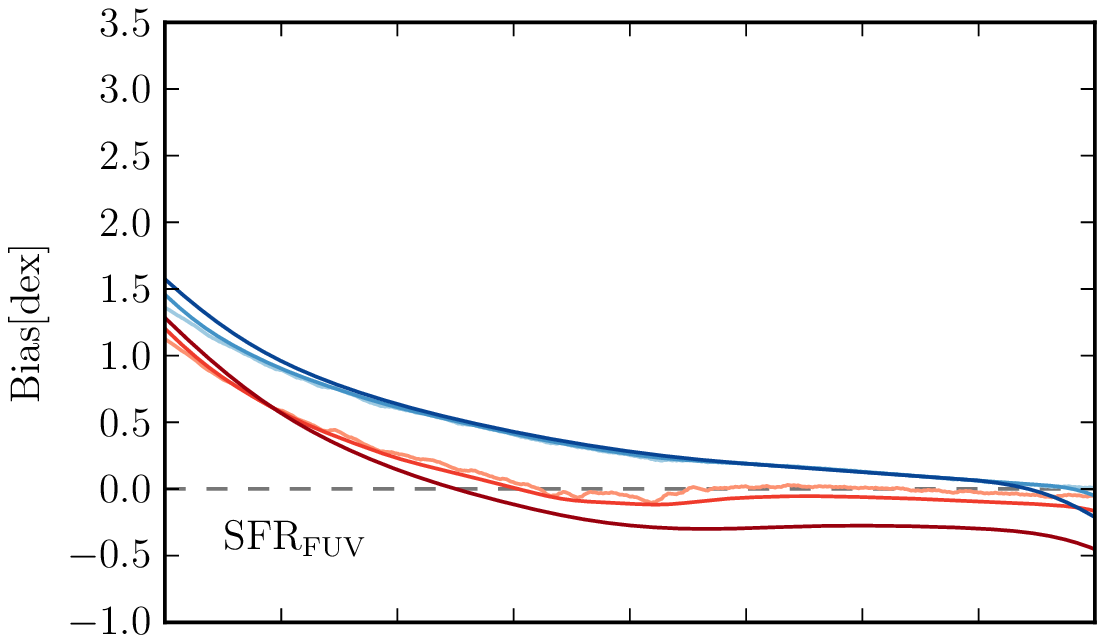} & 
\includegraphics[scale=0.75]{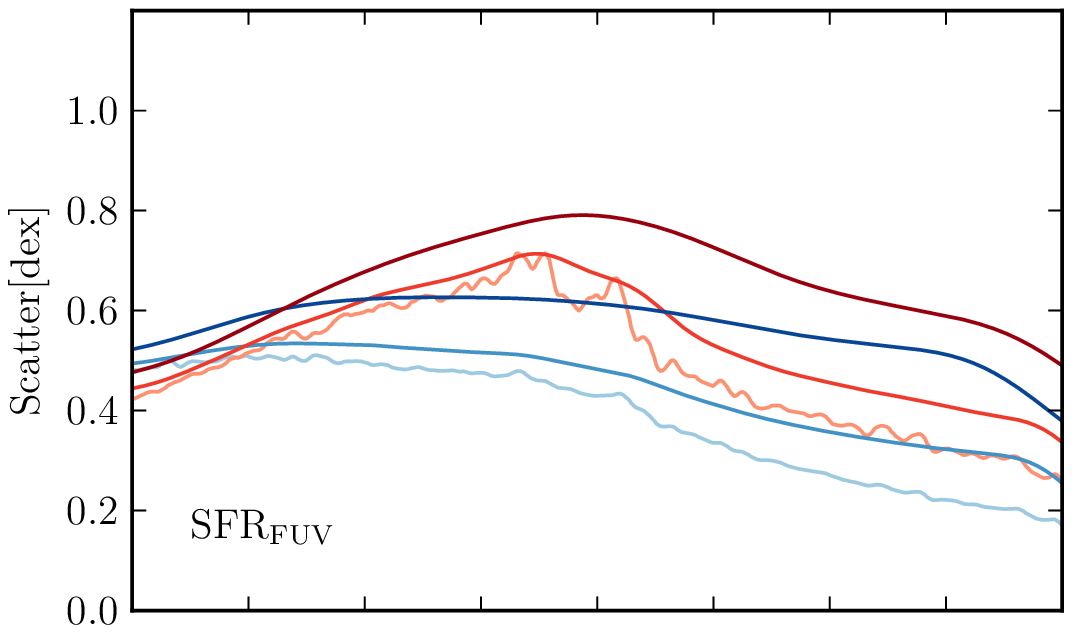} \\
\includegraphics[scale=0.75]{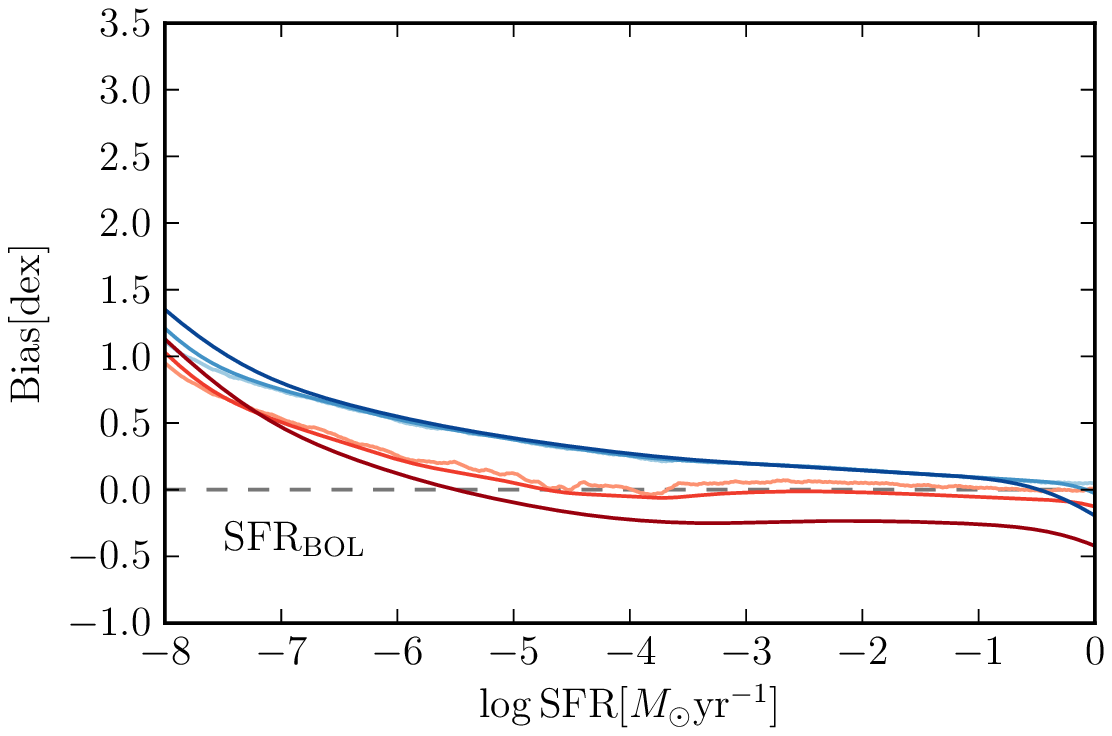} & 
\includegraphics[scale=0.75]{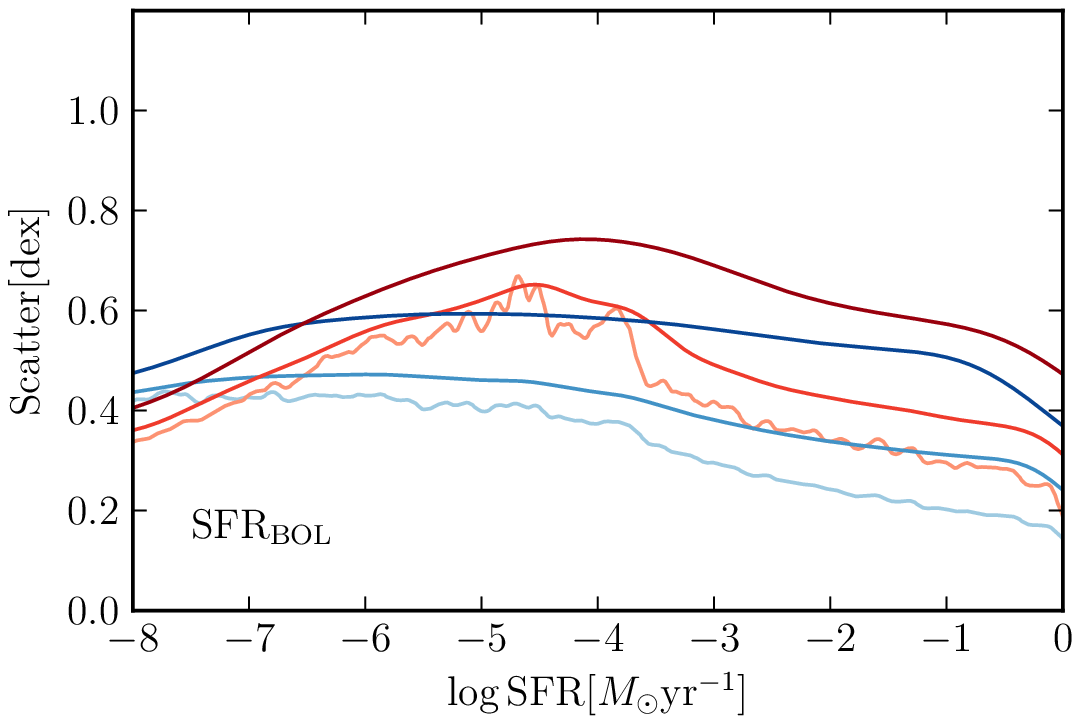} 
\end{tabular}
\caption{Bias (Eq. \ref{eqt:bias}) and scatter (Eq. \ref{eqt:scatter}) due to stochasticity in SFR estimates using the SFIs indicated in each panel.
The lower observational error models produce noisier curves because they are averaging over fewer \slug \ models.}\label{fig:fitplot}
\end{figure*}

Figure \ref{fig:fitplot} shows the bias and scatter as a function of the observed luminosity of the three SFIs we consider in this paper, ionizing/H$\alpha$ luminosity, FUV luminosity, and bolometric luminosity. As expected, we see that both the bias and scatter are reduced at high SFRs, and that both are largest for ionizing luminosity-based SFRs, since they are the most sensitive to the most massive stars. Although it is not immediately apparent from the figure, ionization-based SFIs also have the longest tails (this produces the high value of the bias). We also see the choice of prior has a larger effect in the higher uncertainty observations. This is because there is a bigger dynamic range for the PDF to affect the result. As is always the case, the closer the PDF is to a $\delta$ function, the less a prior matters.

We also see that the uncertainty is characteristically largest at $\log {\rm SFR} \approx -4$. Two effects contribute to this peak, which we can understand with simple order-of-magnitude estimates.  First, the luminosity, particularly the ionizing luminosity, is dominated by stars with masses $\gtrsim 20$ $M_\odot$. For our adopted IMF, these stars contribute a fraction $f_N \sim 10^{-2.5}$ by number. The expected number of stars with  masses $\gtrsim 20$ $M_\odot$ present at any given time is $\langle N\rangle = f_N t_{\rm life} (\mbox{SFR} / \langle M\rangle)$, where $\langle M\rangle\sim 0.5$ $M_\odot$ is the mean stellar mass and $t_{\rm life} \sim 4$ Myr is the lifetime of the very massive stars with which we are concerned. Thus a SFR of $\sim 2\times 10^{-4}$ $M_\odot$ yr$^{-1}$ is the value for which the expected number of very massive stars present at any given time transitions from being $\gtrsim 1$ to $\lesssim 1$, and thus represents something of a maximum in the amount of stochastic flickering. This effect has also been discussed by \citet{cervino3} in the context of the ``Lowest Luminosity Limit''.

The second effect is more subtle, and points to a fundamental limitation of our understanding. We adopt a minimum cluster mass of $20$ $M_\odot$, and, as can be seen from Figure \ref{fig:bigiel}, this imposes a minimum SFR $\log {\rm SFR} \sim -8$ corresponding to the lowest star formation possible. SFRs below this value always produce luminosities of zero in our model. However, this means that the range of possible SFRs for a given observed (non-zero) luminosity has a hard lower limit, and this has the effect of limiting the width of the SFR PDF, and thus the scatter, at the very lowest SFRs. Such a hard edge to star formation is obviously artificial, but it does point out the fact that, at very low SFRs, it is not possible to make a good estimate of the scatter without knowing exactly how star formation and stellar clustering work in regimes where the number of star clusters present at any given time is likely to be zero. Without this knowledge, one cannot calculate in logarithmic space the probability that a galaxy with a SFR of, say, $10^{-5}$ $M_\odot$ yr$^{-1}$ based on the point mass estimate is actually a galaxy with a true SFR of $10^{-8}$ $M_\odot$ yr$^{-1}$ that has just formed a single O star and thus has a temporarily boosted luminosity.

A much more subtle version of this effect, is responsible for the very slight turn-down in bias and scatter that we observe as the SFR approaches 1 $M_\odot$ yr$^{-1}$. For reasons of numerical cost we have not been able to run models with $\log {\rm SFR} \gtrsim 0.3$, and this slightly limits the bias and scatter at the highest SFRs we explore. As is apparent from Figure \ref{fig:fitplot}, however, the effect is very minor.

\subsection{Publicly-Available Tools}

We caution that, while the summary statistics discussed in the previous section are useful rules of thumb, those attempting a proper statistical analysis of their data should make use of the full PDFs and calculate posterior probability distributions from Equation \ref{eqt:sfrpdffinal}. To facilitate such computations, we have made two tools publicly-available at \vizloc\footnote{For different ways of accessing these data products, please contact M. Fumagalli or M. Krumholz.}.

\begin{figure*}
\includegraphics[scale=0.5]{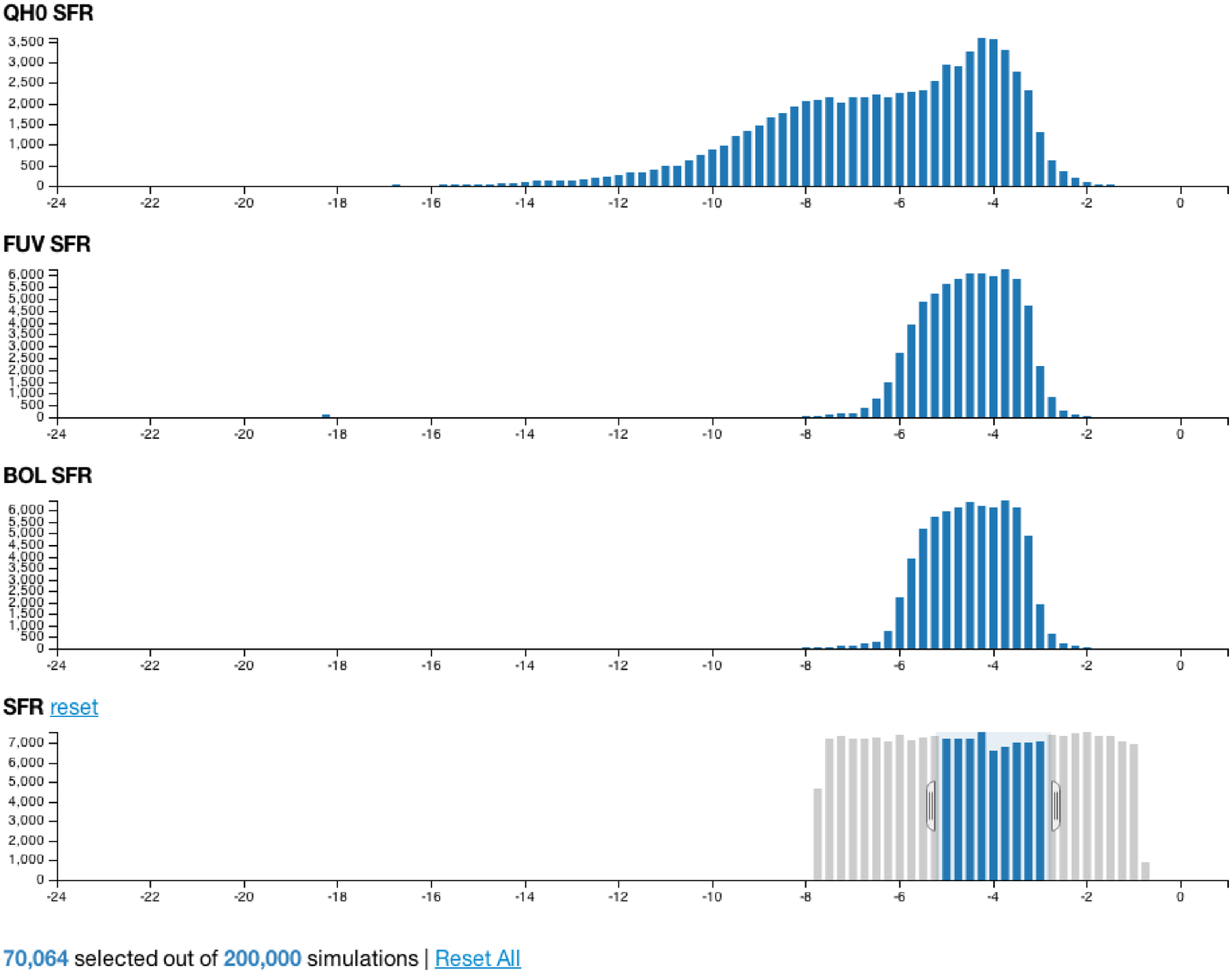}
\caption{Screenshot of interactive data visualization tool for the 4-dimensional parameter space ($\sfr$, $\sfrqho$, $\sfrfuv$, $\sfrbol$). Available at
\vizloc. Selections can be applied to any dimension(s) to show the effects on the others.}
\label{fig:slugpdfscreen}
\end{figure*}

First, we have created an interactive visualization tool; Figure \ref{fig:slugpdfscreen} shows a screenshot. Its operation is as follows. As discussed above, one may think of our simulations as populating a four-dimensional parameter space ($\sfr$, $\sfrqho$, $\sfrfuv$, $\sfrbol$). Either an input theoretical $\sfr$, or an observation of one or more of the star formation tracers, picks out a particular part of this parameter space, and therefore restricts the range of values available for the other tracers. The visualization tool allows users to see these effects by selecting a range of values in one more more of the four parameters. The tool then shows the corresponding range in the other parameters. For example, in the screen shot shown in Figure \ref{fig:slugpdfscreen}, a user has selected a range of intrinsic SFRs centered around $\log\mbox{ SFR} = -4$ (bottom panel), and the tools is displaying the corresponding range of values for $\sfrqho$, $\sfrfuv$, and $\sfrbol$ (top three panels). Versions of the tool are available for both flat and Schechter function priors, and for different clustering fractions (see Section \ref{ssec:sensitivity}).

Second, we have made available both the full output of the \slug~simulations and a set of python scripts to parse them and use them to evaluate Equation \ref{eqt:sfrpdffinal} for a specified set of observational constraints. The basic strategy implemented in the code for calculating $p(\sfr\mid {\rm data})$ is
 \begin{enumerate}
 \item Run the script that loads in the 1.8 million galaxy simulations and performs the kernel density estimate.
 \item Evaluate the density on a grid of SFI values, weighted by the appropriate prior.
 \item Weigh each point in the above grid by the input observational PDF, $p(\vl\mid {\rm data})$. As an example, the posted python code demonstrates how to do this for a Gaussian error bar.
\end{enumerate}
The output is a PDF similar to the one plotted in Figure \ref{fig:sfrpdf}. The entire operation should take a few minutes at most, with most of the time spent in step 1, which only needs to occur once for evaluation of an entire dataset. We note that one of the benefits of our approach, and our code, is that we can easily extend to considering the distribution of SFR given a joint set of constraints. Nothing changes in the formalism since we have thus far always been treating $\vl$ as a vector.

\section{Discussion}\label{sec:discussion}

Having discussed at length the quantitative implications of stochasticity for the interpretation of SFIs, in this section we step back and consider some of the broader implications of our results. We also discuss some caveats and cautions.

\subsection{Star Formation Rate Distributions and the Cosmic Star Formation Rate Budget}

We have already alluded to one important implication of our results: because there is both a systematic bias and a scatter in SFR determinations, and because both of these quantities depend systematically on the observed value for the SFI, there is likely to be a similar systematic bias in observational determinations of the distribution of SFRs in a galaxy population derived using point-mass calibrations. A number of authors have published such determinations based on a variety of SFIs in both the local and high-redshift Universes (to name but a few of many examples, \citealt{salim07a}: FUV at $z\sim 0$; \citealt{bothwell}: FUV plus infrared / bolometric at $z\sim 0$; \citealt{fontanot12a}: FUV plus infrared / bolometric at $z\sim 0.4-1.2$; \citealt{ly12a}: H$\alpha$ at $z\sim 0.5$; \citealt{smit12a}: FUV at $z\sim 4-7$; \citealt{bauer13a}: H$\alpha$ at $z\sim 0.05-0.3$). Our findings suggest that the results of these surveys may suffer from significant systematic errors, with the extent of the problem depending on the tracer used and on the range of SFR being studied. In particular, faint end slopes may need to be revised, as our results open up the possibility that there may be a non-negligible population of galaxies that have significant SFRs averaged over time, but that are missed in observational surveys simply because they happen to have relatively low UV or ionizing photon luminosities at the instant that the observation is made.

We note that, in setting the prior probability distribution used in our Bayesian analysis, we have relied on these potentially flawed measurements.\footnote{This is not a deficiency of our method compared to others, as any non-trivial statistical analysis requires the use of some prior distribution for the SFR, either explicitly or implicitly.} In principle the proper way to address this issue is via forward modeling. Given a parameterized functional form for the SFR distribution (e.g., a Schechter function), one could use $p(\vl\mid\log\sfr)$ to calculate the observed SFI luminosity distribution that would be expected for a particular choice of parameters describing the SFR, and then adjust those parameters iteratively until the predicted SFI luminosity distribution matches the observed one. However, such an approach is beyond the scope of this work, as an accurate forward model would need to be constructed on a survey-by-survey basis, as it would have to fold in uncertainties and errors arising from finite instrumental sensitivity, the color or other cuts used to define the sample, and similar effects.

This issue may also affect determinations of the cosmic SFR budget \citep[e.g.][]{hopkins06a}. These measurements are somewhat less vulnerable to stochasticity than measurements of the SFR distribution, as they necessarily involve averaging over a large number of galaxies and thus averaging out stochasticity (though given the large scatter, the required number of galaxies may be large). If one could in fact observe every H$\alpha$ photon, for example, emitted in a particular field in a given redshift range, there would be no error from stochasticity as long as the field were large enough to have a bulk SFR larger than $\sim 1$ $M_\odot$ yr$^{-1}$. However, in practice measurements of the SFR budget are based on flux-limited samples, and stochasticity can interact with the flux limit by scattering some galaxies with low average SFRs into the sample, while scattering others with higher SFRs out of it. Which of these two effects dominates is a subtle question, since there are more low-luminosity galaxies that could potentially scatter above the flux cut, but the skewness of the PDF is such that galaxies are more likely to be under- than over-luminous for their SFR. Again, rigorous treatment of this issue requires that the study's selection function be analyzed properly with Monte Carlo simulations.

\subsection{Kennicutt-Schmidt Relations}

Another area where luminosity-dependent bias and scatter in SFIs can cause problems is in empirical determinations of the relationship between gas and star formation in galaxies, generically known as Kennicutt-Schmidt relations \citep{schmidt59a, ken98}. Prior to the past decade, such relationships were generally measured as integrated quantities over fairly large spiral galaxies. In the past decade, however, there has been a concerted effort to push these measurements to galaxies with lower global SFRs \citep[e.g.,][]{lee2009, boselli, meurer}, and to ever-smaller spatial scales within large galaxies \citep[e.g.,][]{wong,kennicutt07a,miki08,bigiel,bigiel10,schruba10a,onodera10a,bolatto11a, calzetti12a, momose13a, leroy2013}. These efforts have pushed the data into realms of ever-lower absolute SFR, and thus greater vulnerability to stochasticity \citep{Kruijssen2014}.

To take one example, for the lowest gas surface density bin in the sample of \citet{bigiel10}, the median SFR surface density is inferred to be a bit over $10^{-6}$ $M_\odot$ yr$^{-1}$ kpc$^{-2}$. For the mean pixel size of 600 pc used in the study, this corresponds to $<10^{-6}$ $M_\odot$ yr$^{-1}$. The study uses FUV as its SFI of choice, and consulting Figure \ref{fig:fitplot}, we see that, for a Schechter function prior and assuming negligible observational errors, we expect a scatter of $\sim 0.5$ dex from stochasticity alone. If we adopt a flat prior distribution of SFRs (perhaps reasonable inside a galaxy), we also expect a similar amount of bias. This will obviously affect the mean relation that one infers between gas and SFR, and it should be accounted for when fitting the observations. Qualitatively, the net effect of stochasticity is likely to be that the inferred relationship between SFR and gas surface density is too steep at the lowest SFRs (due to the bias) and that the inferred scatter will be larger than the true one (due to the extra scatter in the SFI-SFR relation imposed by the stochasticity).

\begin{figure*}
\begin{tabular}{cc}
\includegraphics[scale=0.7]{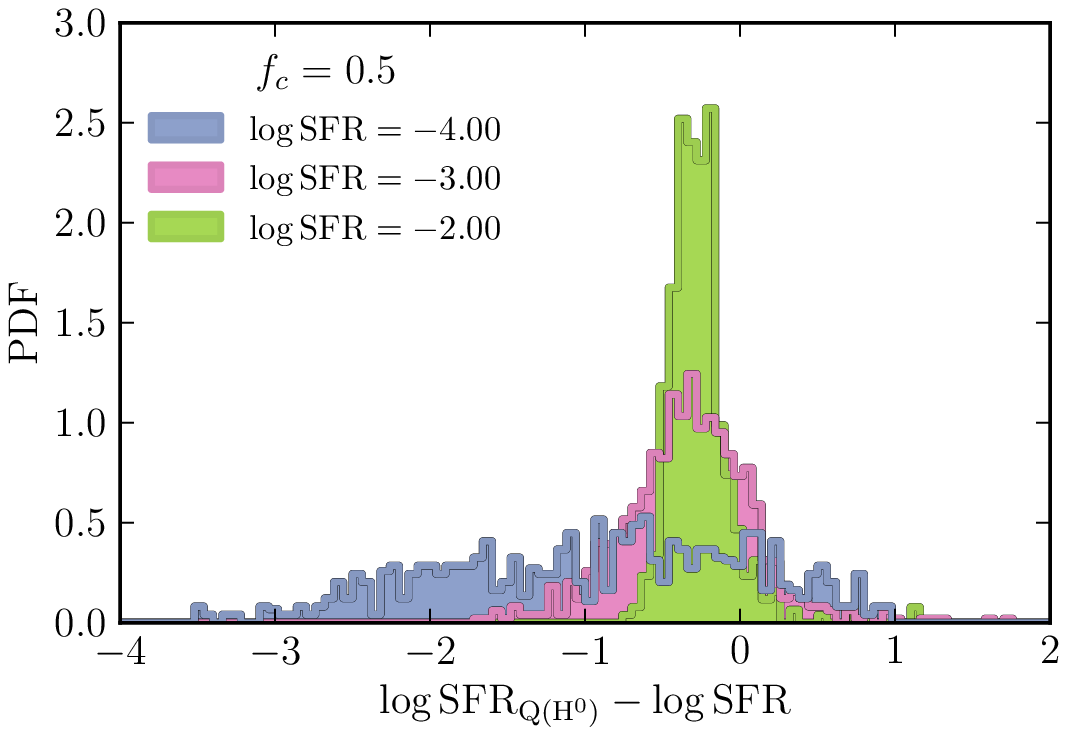}&
\includegraphics[scale=0.7]{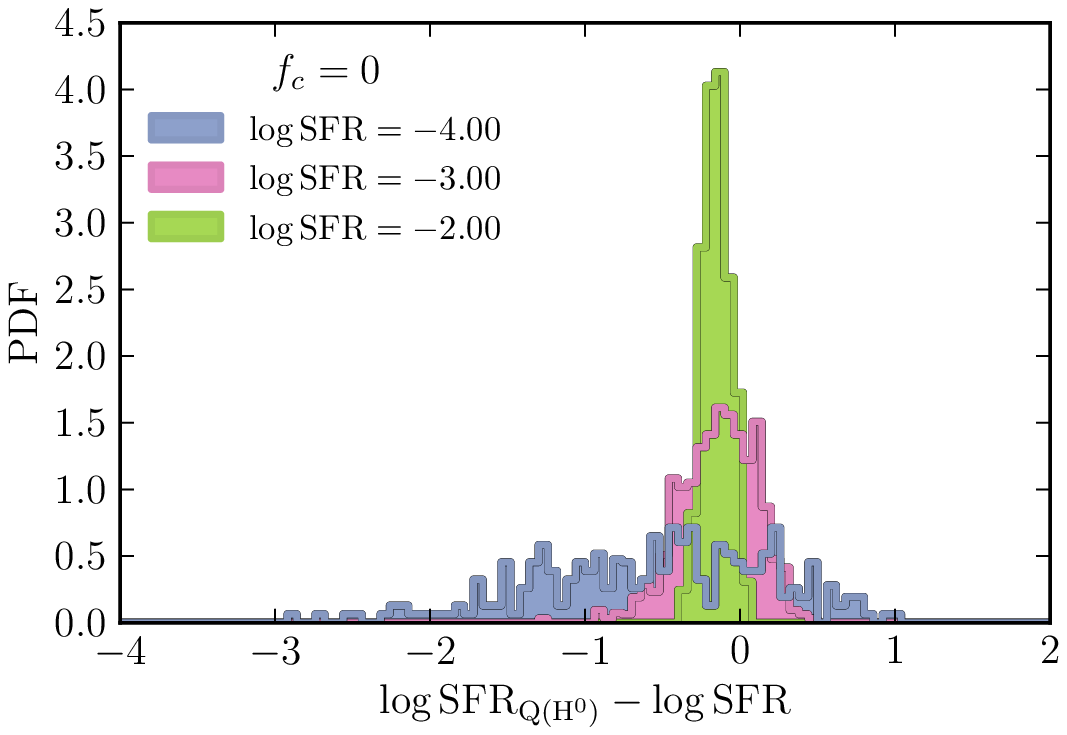}\\
\includegraphics[scale=0.7]{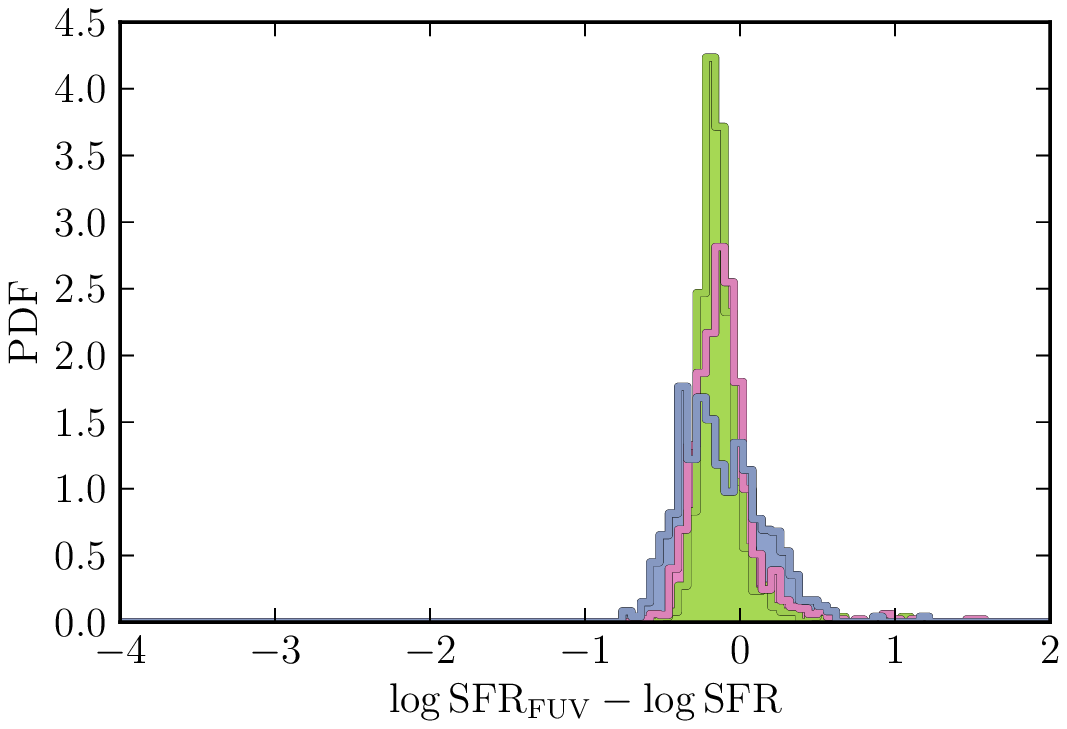}&
\includegraphics[scale=0.7]{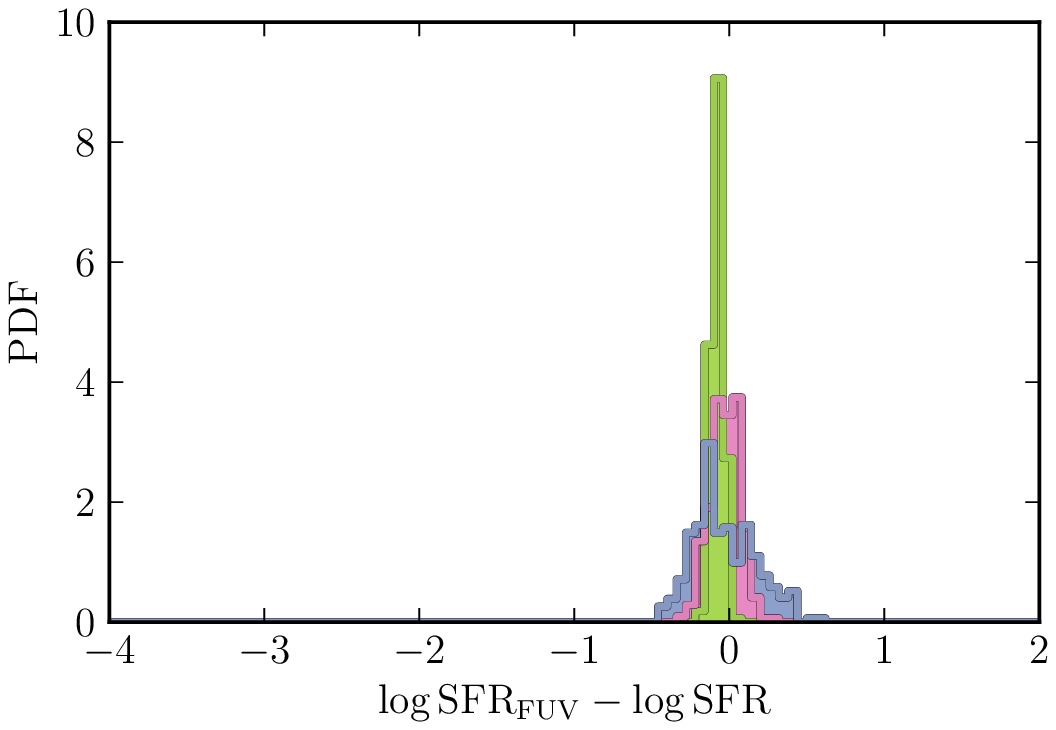}\\
\includegraphics[scale=0.7]{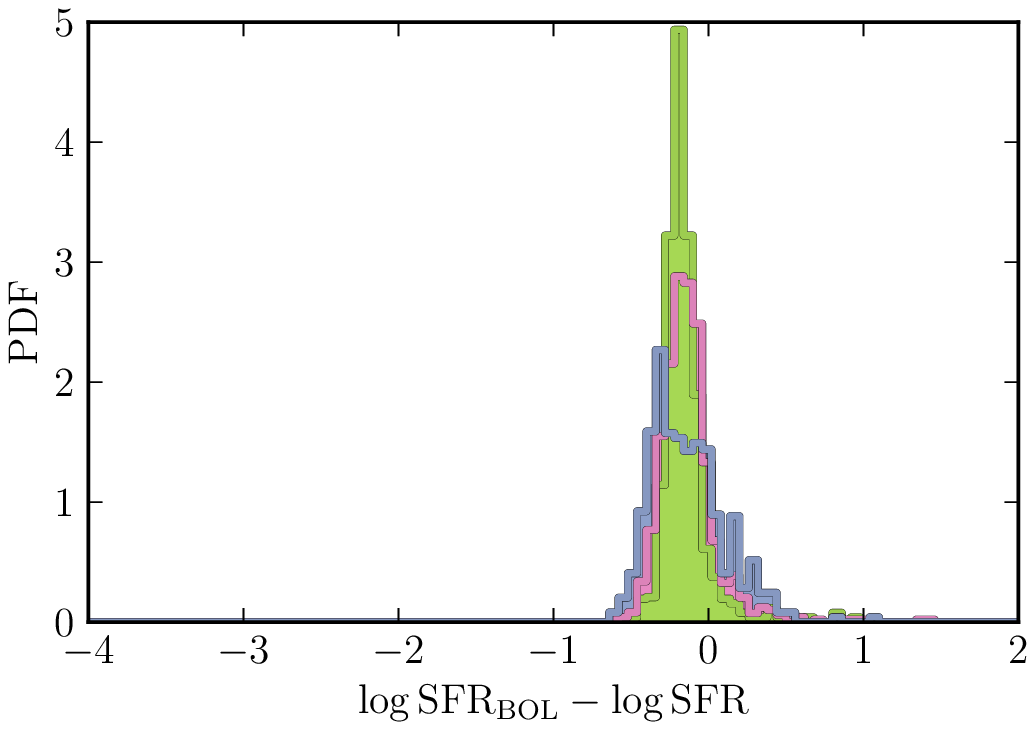}&
\includegraphics[scale=0.7]{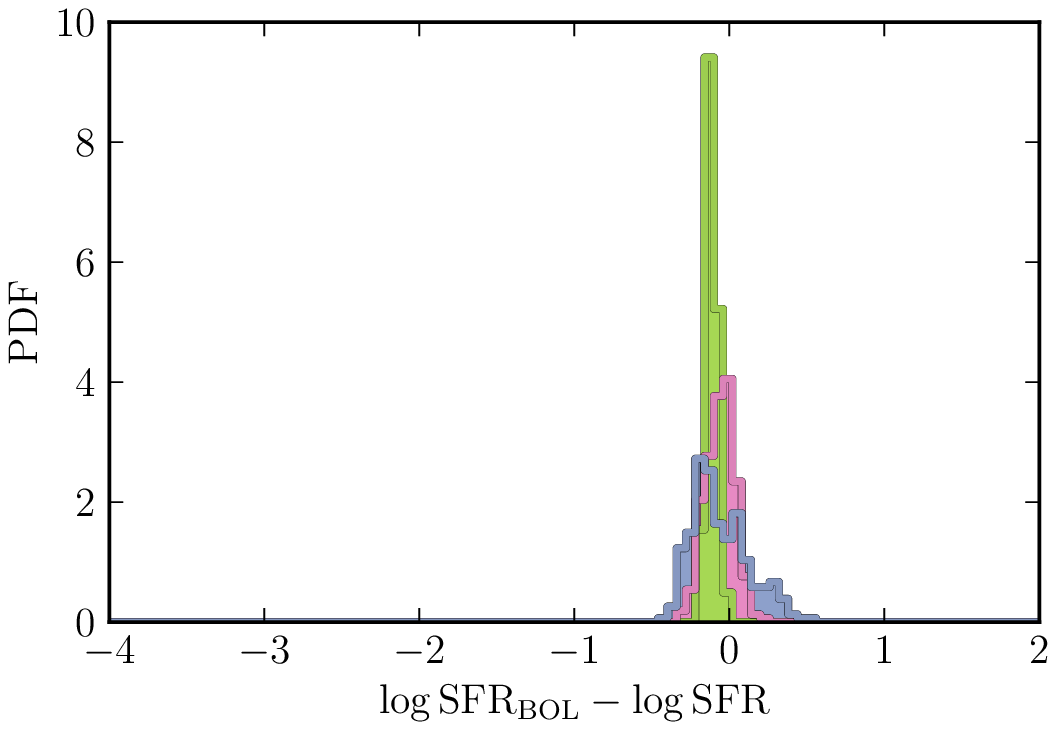}
\end{tabular}
\caption{ Same as Figure \ref{fig:sfi_pdfs} but for $f_c=0.5$ (left) and $f_c=0$ (right).}
\label{fig:sfi_pdfs2}
\end{figure*}

\subsection{Sensitivity to Parameter Choices}
\label{ssec:sensitivity}

We end this discussion with a caution regarding the sensitivity of our results to some of the parameters we have chosen in our \slug~simulations. The results obviously depend to some extent on the choice of stellar evolutionary tracks and atmosphere models, but this is true even in the absence of stochasticity. The parameters that are unique to our stochastic models are those that describe how stars are clustered.
A full analysis of the effects of varying the cluster
mass function's minimum and maximum mass, as well as its power law index and the total fraction of stars formed in clusters, is well beyond the scope of this paper.
However, to explore the effects of clustering to gain some intuition, we focus on a single parameter: the total fraction of stars formed in clusters $f_c$.\footnote{An important note on nomenclature: some authors whose interest lies primarily in stellar dynamics \citep[e.g.,][]{portegies-zwart10a} limit the definition of star clusters to include only those stellar structures that are gravitationally-bound and dynamically-relaxed. These are distinguished from associations -- collections of stars that are born in spatial and temporal proximity to one another, but need not be bound or relaxed. Since we care only about the temporal correlation of star formation, and not about the dynamical evolution of the structures in which the stars form, we are interested in a much more expansive definition of clustering, one that includes both clusters and associations. Thus our $f_c$ parameter is not directly comparable to the parameter $\Gamma$ that is sometimes introduced to denote the fraction of star formation that occurs in structures the remain bound after the transition from gas-dominated to gas-free evolution \citep[e.g.,][]{bastian08a}.} This is likely the single most important parameter. Our default choice is $f_c=1$. This is motivated by the observation that, in the Milky Way, most star formation occurs in clusters \citep{ladalada}, and by the result that models with $f_c=1$ provide an excellent match to the observed distribution of H$\alpha$ to FUV ratios in local dwarf galaxies \citep{mikiletter}. However, to investigate how our results would change if we alter this parameter, we run roughly $15,000$ unclustered models ($f_c=0$) and $25,000$ with $f_c=0.5$. These models are uniformly distributed in $\log \sfr$ between $-4$ and $-2$.
 
Figure \ref{fig:sfi_pdfs2} shows the PDFs of offset between SFI and true SFR that we obtain from the unclustered and reduced clustering runs; it should be compared with Figure \ref{fig:sfi_pdfs} for our fiducial case. The comparison indicates that reducing the clustering can significantly reduce the spread of SFI values produced at fixed SFR. This will correspondingly significantly decrease the scatter in the inferred SFR PDFs.
 
This result implies that, at least at low SFRs, it is crucial to understand the clustering properties of star formation in order to do something as simple as inferring a SFR. A more accurate determination of stellar clustering parameters, and whether they vary with galactic environment, is therefore urgently needed. Our fiducial parameters are reasonable first approximations based on empirical constraints from local galaxies, but if clustering parameters vary systematically with galaxy properties, the effects of stochasticity on inferences of the SFR may as well.

\section{Summary}\label{sec:summary}
While star formation in galaxies is often imagined as a continuous, ongoing process, observations tell us that the actual distribution of stellar ages is highly stochastic, with stars mostly forming in discrete bursts of finite size. At sufficiently high star formation rates (SFRs), the overall process of star formation in a galaxy consists of many such bursts, and the continuous approximation is reasonable. In this paper, we use the Stochastically Lighting Up Galaxies (\textsc{slug}) code to investigate what happens at lower SFRs when this approximation begins to break down, with particular attention to how this breakdown affects our ability to infer the underling SFR using a variety of star formation indicators (SFIs).

We show that the generic effect of stochasticity is to produce a broad probability distribution function (PDF) for SFI luminosity a fixed SFR. The breadth of the PDF depends on both the SFI being used and on the true SFR. We then devote the bulk of the paper to understanding the implications of this spread in SFI at fixed SFR for the inverse problem of inferring the true SFR given an observed SFI. We derive an analytic expression for the PDF of true SFR given a set of observational constraints, and provide software to evaluate this PDF using our simulation results and a set of user-specified observational constraints.

Using this formalism, we show that the process of inferring the SFR from an observed SFI is subject to scatter, and, more worryingly bias, meaning that the process of simply converting between SFI and SFR using the standard calibrations that apply at higher SFRs is likely to lead to systematic errors when used at low SFRs. The strength of the bias and scatter depend on both the observed values of the SFI and on its observational uncertainty, and on the choice of SFI. Ionization-based SFIs such as H$\alpha$ emission in particularly can be problematic due to the very short timescales over which they average; for such indicators, a scatter of several tenths of a dex is expected even at inferred SFRs as high as $\sim 1$ $M_\odot$ yr$^{-1}$. Even for indicators much less subject to scatter such as FUV luminosity, for measurements with non-trivial observational uncertainty, biases of up to $\sim 0.5$ dex are possible.

Finally, we discuss the implications of these results for efforts to construct ``luminosity functions" of SFR, for estimates of the cosmic SFR budget, and for inferences of the Kennicutt-Schmidt Law relating gas content to SFR. The Legacy Extragalactic UV Survey (LEGUS; Calzetti et al., 2014, in preparation) will provide a valuable data set for this type of analysis.

\section*{Acknowledgments}
We thank the referee, Richard de Grijs, for useful comments that have improved this manuscript.
R.L.dS. and M.R.K. acknowledge support from NASA through Hubble Award Numbers HST-AR-13256.01-A and HST-GO-13364.26-A issued by the Space Telescope Science Institute, which is operated by the Association of Universities for Research in Astronomy, Inc., under NASA contract NAS 5-26555. The work of R.L.dS was supported by the National Science Foundation Graduate Research Fellowship. M.R.K. acknowledges support from NSF grant AST-0955300, and from NASA ATP grant NNX13AB84G. MF acknowledges support by the Science and Technology Facilities Council [grant number  ST/L00075X/1]. To access the simulations presented in this work, please contact the authors. 
 
\bibliographystyle{mn2e}
\bibliography{slugrefs}

\label{lastpage}

\end{document}